\documentclass[12pt,preprint]{aastex}

\shorttitle{GOODS-CDFS radio sources}
\shortauthors{Afonso et al.}

\begin{document}

\title{Optical and X-ray Identification of Faint Radio Sources in the
GOODS-S ACS Field}

\author{J. Afonso\altaffilmark{1,2}, B. Mobasher\altaffilmark{3,4},
A. Koekemoer\altaffilmark{3}, R. P. Norris\altaffilmark{5}, 
L. Cram\altaffilmark{6}}

\altaffiltext{1}{Universidade de Lisboa, Faculdade de Ci\^{e}ncias,
Observat\'{o}rio Astron\'{o}mico de Lisboa, Tapada da Ajuda, 1349-018
Lisboa, Portugal; jafonso@oal.ul.pt} \altaffiltext{2}{Centro de Astronomia e
Astrof\'{\i}sica da Universidade de Lisboa} \altaffiltext{3}{Space
Telescope Science Institute, 3700 San Martin Drive, Baltimore MD
21218, USA} \altaffiltext{4}{Also affiliated to the Space Sciences
Department of the European Space Agency} \altaffiltext{5}{Australia
Telescope National Facility, PO Box 76, NSW 1710, Epping, Australia}
\altaffiltext{6}{The Australian National University, Canberra ACT 0200 
Australia}
\begin{abstract}

We present optical and X-ray identifications for the sixty-four radio sources
in the GOODS-S ACS field revealed in the ATCA 1.4\,GHz survey of the
Chandra Deep Field South. Optical identifications are made using
the ACS images and catalogs, while the X-ray view is provided by the 
Chandra X-ray Observatory 1\,Ms observations. Redshifts for the identified
sources are drawn from publicly available catalogs of spectroscopic 
observations and multi-band photometric-based estimates. Using this
multiwavelength information we provide a first characterization of the
faint radio source population in this region. The sample contains a mixture 
of star-forming galaxies and active galactic nuclei, as identified by
their X-ray properties and optical spectroscopy. A large number of 
morphologically disturbed galaxies is found, possibly related to the
star-formation phenomena. In spite of the very deep optical
data available in this field, seven of the sixty-four radio sources have
no optical identification to $z_{850}\sim 28$\,mag. Only one of these
is identified in the X-rays.

\end{abstract}

\section{INTRODUCTION}

The Great Observatories Origins Deep Survey (GOODS)
\citep{Giavalisco04} provides deep multi-wavelength observations in
the Hubble Deep Field-North (HDF-N) and Chandra Deep Field-South
(CDFS).  The fields have been imaged by the {\it Hubble Space
Telescope} (HST), using the Advanced Camera for Surveys (ACS) in four
bands (F435W, F606W, F775W and F850LP).  At infrared wavelengths,
observations by the {\it Spitzer Space Telescope} use the IRAC
(3.6-8\,$\mu$m) and MIPS (24\,$\mu$m) instruments.  The deepest X-ray
surveys currently available have been made in the GOODS fields, using
the {\it Chandra X-ray Observatory} and {\it XMM-Newton}.  The GOODS
fields are also targets of many ground-based studies using 4- and
8-meter class telescopes, providing imaging and spectroscopic data in
optical and near-infrared bands. The richness and depth of data
implies that GOODS will be a foundation for the study of galaxy
formation and evolution over the next few years.

Space and ground-based observations and data acquired up to 2004 are
described by \citet{Giavalisco04}. The available spectroscopic
redshifts for these fields, combined with the multi-waveband imaging
dataset, have been used to calibrate and determine photometric
redshifts of many sources to faint magnitudes
\citep[e.g.][]{Mobasher04,Wolf04}.

Deep radio observations of GOODS represent an excellent opportunity to
explore a number of unresolved questions regarding radio emission from
galaxies, since GOODS sources have a plethora of multiwavelength data
available for their characterization. Additionally,
radio data on GOODS sources can, in some situations, resolve
uncertainties and/or provide unique insights about the astrophysical
processes at play.

It is believed that the decimetric radio emission from galaxies is
dominated by the synchrotron emission from electrons accelerated
either by shock waves associated with supernovae (and hence star
formation, SF), by processes energized by a massive nuclear object
(an active galactic nucleus, AGN), or by a combination of the
two \citep{Condon92}.

Radio emission associated with AGNs is responsible for the great
majority of decimetric sources brighter than about 
$\approx 5$\,mJy. The optical hosts of these sources include massive elliptical
galaxies, QSOs and Seyfert galaxies. Radio selection of sources
brighter than a few mJy preferentially selects for ``radio-loud'' AGNs
having relatively high radio-to-optical luminosity ratios. AGNs also
exist in the sub-mJy population: some are distant radio-loud sources,
and others are intrinsically low-luminosity radio sources hosted by
objects similar to those hosting radio-loud AGNs. Observations of the
radio emission from AGNs provide information about the existence and
properties of central black holes, especially when combined with
observations of X-ray power \citep[e.g.][]{Merloni03}, as well as clues
to some properties of the interstellar and intergalactic medium.

Among samples of radio sources selected at ever-fainter sub-mJy flux
densities, the proportion of AGNs systematically declines in favor of
sources energized by star formation 
\citep[e.g.][]{Windhorst85,Georgakakis99,Afonso05}.  The
optical hosts of these starforming radio sources 
include blue galaxies with optical colors revealing a young stellar
population, and extremely red galaxies with evidence for a very high
degree of dust extinction. High resolution optical imaging often 
reveals disturbed morphologies indicative of 
interactions and/or merging activity among the sub-mJy 
radio population \citep[e.g.][]{Richards98}.

There is a tight correlation between the decimetric radio luminosity
and the far-infrared (FIR) luminosity of starforming galaxies. In such sources
the FIR emission is believed to be thermal emission from dust heated
by stellar UV and optical emission. When the extinction is so high
that almost all of the stellar luminosity is absorbed by dust, the FIR
luminosity measures the total UV/optical power and hence (if the
initial mass function is known) the rate of star formation
\citep[e.g.][]{Yun01}. The correlation between FIR and radio luminosity
allows the use of radio power observations to determine star-formation
rates for galaxies that are too faint to be detected as FIR
sources. Importantly, decimetric radio emission is free of extinction.

Deep decimetric observations of GOODS fields thus offer
astrophysically useful data on black hole properties, star-formation
processes, properties of the interstellar and intergalactic medium,
and the relative densities and evolution of some major galaxy
populations. Accordingly, this paper commences the investigation of
radio sources in the GOODS CDFS field, in particular the HST-ACS
region which has homogeneous optical, radio and X-ray coverage. To
assist building a resource for the entire community, optical and X-ray
identifications of the sources revealed in the recent ATCA 1.4\,GHz
radio observations (Koekemoer et al. 2006, in preparation) are reported 
promptly.
Throughout this paper, and unless otherwise noted, we adopt $H_0 =
70\,$ km\,s$^{-1}$\,Mpc$^{-1}$, $\Omega_M = 0.3$, and $\Omega_\Lambda
= 0.7$.

\section{Multiwavelength observations}

\subsection{Radio observations}

The CDFS was observed at radio (1.4\,GHz) wavelengths using the
Australia Telescope Compact Array (ATCA). A total of 120 hours of
observations were obtained, resulting in a 1.2\,square degree area
field covering the CDFS to a limiting (1\,$\sigma$) sensitivity of
$\approx 14$\,$\mu$Jy and a beam size of $16.8\arcsec \times
6.95\arcsec$. Source detection provided a catalog of 683 sources,
with integrated flux densities between 61\,$\mu$Jy and 170\,mJy.
Koekemoer et al. (2006, in preparation) present the observations, 
data reduction methods and the approach used to construct the radio 
source catalog.  

\begin{figure}
\scalebox{1.0}{
\rotatebox{0}{
\includegraphics{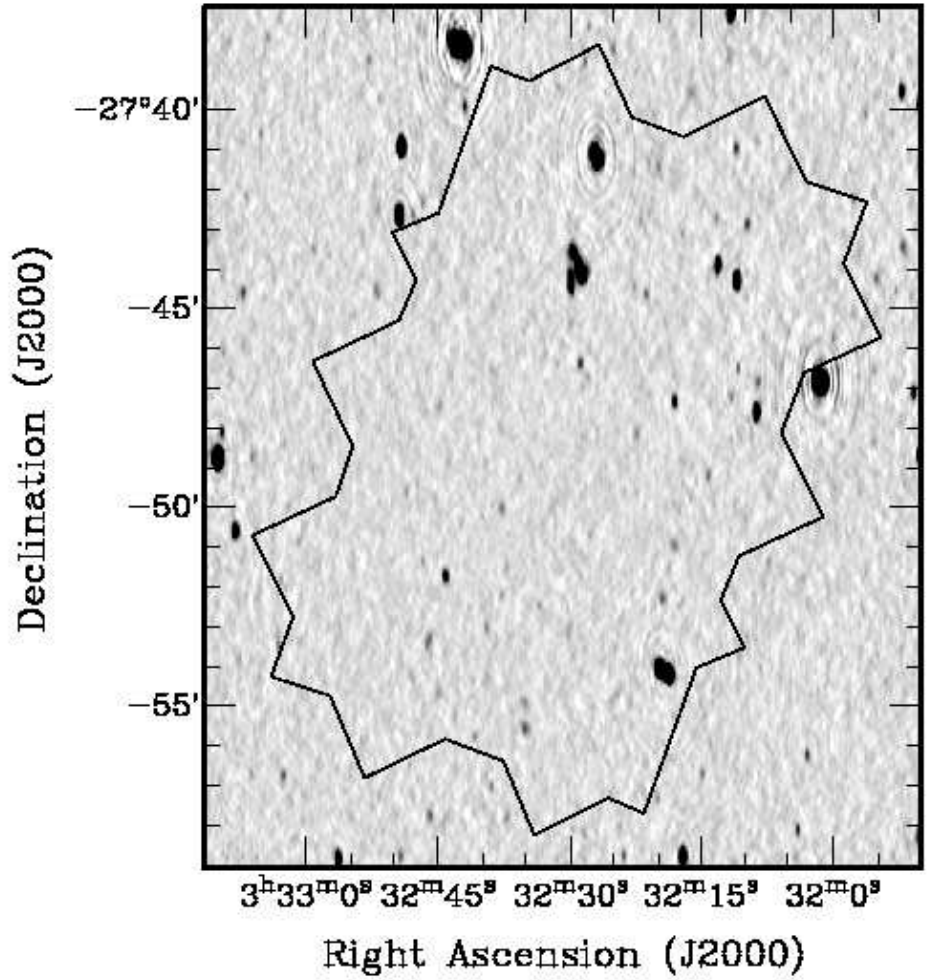}
}}
\caption{The ATCA radio 1.4\,GHz image of the GOODS-S ACS field (outlined). 
The 64 radio sources detected within this field have integrated 
1.4\,GHz flux densities ranging from 63\,$\mu$Jy to 20\,mJy. 
\label{fig:radioview}}
\end{figure}

Within the GOODS ACS CDFS region, a total of 64 radio sources are
found, with 1.4\,GHz flux densities between 63\,$\mu$Jy and 20\,mJy.
Figure~\ref{fig:radioview} shows the radio view of the this field.
This sample and its optical and X-ray properties are presented in 
the present paper.

\subsection{Optical observations}

The HST-ACS observations of the CDFS taken as part of the GOODS
project \citep{Giavalisco04} were used. The data release v1.0
includes $F435W$ ($B_{435}$), $F606W$ ($V_{606}$), $F775W$ ($i_{775}$) and
$F850LP$ ($z_{850}$) reduced, calibrated, stacked and mosaiced images and 
corresponding catalogs, providing photometry of sources detected in the
$z_{850}$-band. Combined $B_{435} + V_{606} + i_{775} + z_{850}$ images were
made, in order to search for faint optical counterparts of radio
sources not detected in the single $z_{850}$ band.

\subsection{X-ray observations}

The deep Chandra X-ray Observatory observations of the CDFS
\citep{Giacconi02} were used. The integration time amounts to 1\,Ms,
being one of the deepest X-ray observations ever taken.  

Initially, the X-ray point-source 
catalogs from \citet{Alexander03} were searched for counterparts of
radio sources. The ${\rm S/N=3}$ sensitivity limits in the 
$\approx 1$\,arcmin$^{2}$ region at the aim point are $\approx 5.2\times
10^{-17}$\,erg\,cm$^{-2}$\,s$^{-1}$ and $\approx 2.8\times
10^{-16}$\,erg\,cm$^{-2}$\,s$^{-1}$ in the 0.5--2.0\,keV and
2.0--8.0\,keV bands, respectively \citep{Alexander03}.

The catalogs presented in \citet{Giacconi02} were also inspected for X-ray
detections not reported in \citet{Alexander03}:
the different data reduction and source detection methods result in 
a few faint sources being detected in only one of these works, which is 
nevertheless still relevant for the current study. However, this imposes 
some non-uniformity for the reported X-ray characteristics. While the 
catalogs produced by \citet{Alexander03} provide counts and fluxes 
in various bands covering the 0.5--8.0\,keV interval, and effective photon 
index\footnote{As described in \citep{Alexander03}, the 
effective photon index $\Gamma$ is calculated for a power-law model with 
the Galactic column density, taking into account the ACIS quantum 
efficiency degradation.} ($\Gamma$), those by \citet{Giacconi02} provide only 
counts and fluxes in the 0.5--2.0\,keV and in the 2.0--10.0\,keV bands.

In order to provide comparable quantities for sources detected in these 
different catalogs, we estimate the X-ray flux in the 2--8\,keV band 
for three sources detected only in the \citet{Giacconi02} catalogs, 
using the reported 2.0--10.0\,keV flux and assuming a power-law photon 
index of 1.4, the average value for the remaining X-ray detected sources, 
as mentioned below. Still, the different reduction techniques and assumptions
(considering or not the ACIS degradation, for example) between these X-ray
catalogs should be noted.

\section{Source Identification}

Optical counterparts of radio sources in the GOODS ACS field were identified 
using the likelihood ratio method of \citet{Sutherland92}. For each radio 
source, the optical identification with the highest reliability 
($\mathcal{R}$), if above 20\%, 
was taken as the real optical counterpart - if several 
optical identifications have similar values for $\mathcal{R}$, 
the various possibilities are considered. Identifications were 
inspected visually to check for special situations where the likelihood ratio
method would not apply, as in the case of non-independent sources (either in 
the radio or in the optical).

The identification of the X-ray counterparts was performed by searching the 
3\,$\sigma$ radio position error region. Visual inspection was used to 
associate the optical and X-ray identifications.

A few sources require special mention. 
Source \#13 (ATCDFS\_J033213.28) coincides with two
optical sources in the ACS $z_{850}$ image, although the GOODS
catalog shows a single identification. SExtractor was used to
deblend the two objects and obtain positions and optical magnitudes
for both. Source \#19 (ATCDFS\_J033218.62) and
source \#21 (ATCDFS\_J033219.62) are 12\arcsec~apart and display similar 
radio fluxes (4.8\,mJy and 5.9\,mJy, respectively). They are regarded as 
being the two radio lobes of the same radio galaxy, optically identified 
with a cD-type galaxy located roughly halfway between the two. 
Source \#55 (ATCDFS\_J033244.16) is identified with an extended
interacting pair of galaxies, their optical positions lying just outside 
and beside the radio position error ellipse. This system is considered as 
the optical counterpart of the radio emission.  

Table~\ref{tab:ids} presents the final source identifications. The columns 
display:

(1) Source number;

(2) Radio source name;

(3) 1.4\,GHz flux density and associated error;

(4) ACS identification;

(5) reliability ($\mathcal{R}$) of the ACS identification;

(6) distance in arcsec between the radio and the ACS positions;

(7) ACS $z_{850}$ (AB) magnitude;

(8) Chandra X-ray Observatory identification;

(9) distance in arcsec between the radio and the CXO positions;

(10) X-ray flux in the 0.5 - 2\,keV band;

(11) X-ray flux in the 2 - 8\,keV band;

Radio sources with no optical counterpart show the $z_{850}$ 3\,$\sigma$ 
upper limit, estimated from the pixel rms 
(0.0008-0.0015\,counts\,s$^{-1}$\,pixel$^{-1}$), integrated over a 
$0.2\arcsec \times\ 0.2\arcsec$~aperture ($4 \times\ 4$ ACS pixel box).
The reliability calculation for optical identifications assumes each optical
detection individually. Whenever a merger or interacting pair of 
galaxies is identified as the optical counterpart, and one of the sources
clearly dominates $\mathcal{R}$, we present in Table~\ref{tab:ids} only the
highest value of $\mathcal{R}$, noting that this will be a lower limit for the 
true reliability of the optical system. Sources with an X-ray identification 
from \citet{Giacconi02} are identified by their reported source number in 
that work. 

Of the 64 radio sources in the GOODS ACS CDFS region, two are considered to
be radio lobes of the same radio galaxy. For most of the sample, a reliable 
unique optical counterpart is found, but eight of the radio sources have
several possible optical counterparts: five of these 
(sources \#13, \#47, \#48, \#51 and \#55) arise from likely interacting 
pairs, for which we present in Table~\ref{tab:ids} data for both 
members, while the remaining three (sources \#41, \#52 and \#60) are
ambiguous identifications, each with two seemingly independent likely 
optical counterparts. The X-ray detection for sources \#41 and \#52 
suggests which optical counterpart is the real one. Seven (11\%) radio sources
have no optical counterpart. One of these, source \#42 (ATCDFS\_J033233.44),
is associated with faint X-ray emission reported in the 
\citet{Giacconi02} catalog. 

Figure~\ref{fig:pstamps} shows gray-scale images in
$B_{435} + V_{606} + i_{775} + z_{850}$, with radio contours and X-ray
detections overlayed.

\section{Source characterization}

Using the deep multiwavelength data available, 
we provide a first characterization of the radio source population in 
the GOODS-S ACS field. Several indicators were employed, as described below.

\subsection{X-ray-to-optical flux ratios}

The X-ray-to-optical flux ratio can provide a first 
indication of the nature of a galaxy 
\citep[e.g.,][]{Maccacaro88,Stocke91,Hornschemeier03}. 
Here, we estimate the X-ray-to-optical flux ratio from the relation

\begin{equation}
f_x/f_{opt}=\log (f_{\rm 0.5-2\,keV}) + 0.4\,z_{850} + 6.12
\end{equation}

\noindent where the X-ray flux is measured in erg\,cm$^{-2}$\,s$^{-1}$ in the 
0.5-2\,keV band. This expression results from the X-ray-to-optical flux ratio 
definition of \citet{Stocke91}, that involved 0.3-3.5\,keV flux and $V$-band 
optical magnitude. These quantities were converted to 0.5-2\,keV flux and 
$z_{850}$ magnitude assuming a power-law X-ray spectral energy distribution 
with $\Gamma=1.4$ and a $V-z_{850}$ color of 1.2\,mag,  
which are the averages for galaxies in the current sample with detections 
in these bands. Assuming this expression, AGNs will {\it typically}
display X-ray-to-optical flux ratios above $-1$, while star-forming galaxies 
{\it tend} to display lower values 
\citep[e.g.][]{Stocke91,Schmidt98,Shapley01,Hornschemeier01,Alexander02}. 

\subsection{Radio-to-optical flux ratios}

We also define a radio-to-optical flux ratio as

\begin{equation}
f_{1.4}/f_{opt}=\log ({\rm S}_{\rm 1.4\,GHz}) + 0.4\,z_{850} - 6.56
\end{equation}

\noindent where we use the measured radio flux density at 1.4\,GHz in mJy.
Although this is not an AGN/star-formation discriminator, since high 
radio-to-optical fluxes can be seen in both radio-powerful AGN or 
dust-rich galaxies, it is still useful to identify extreme sources in 
the sample.

\subsection{Spectroscopic Redshifts}

The CDFS is being intensively studied by several groups. In
particular, spectroscopic observations of the optical counterparts of
many X-ray sources have been made. 
The counterparts of the faint radio sources
in the GOODS ACS CDFS field presented here were thus searched for the
availability of spectroscopic redshifts from
\citet{Szokoly04}, \citet{Lefevre04}, \citet{Vanzella04} and \citet{Mignoli05}.
The first reports spectroscopic observations of possible optical 
counterparts of X-ray sources in the CDFS, down to $R\sim 26$\,mag, using the
VLT with the FORS1/FORS2 spectrographs. \citet{Lefevre04} 
describe the results of the VIMOS VLT Deep Survey around the CDFS, 
which includes 784 redshifts in the ACS area, down to $I_{AB}=24$\,mag.
\citet{Vanzella04} present the VLT-FORS2 spectroscopic observations 
of sources in the GOODS-S ACS region. Source selection in that investigation 
was optimized to detect high redshift objects, in order to take advantage 
of the characteristics of FORS2 (high throughput and spectral resolution, and 
reduced fringing at red wavelengths), mostly down to $z_{850}=24.5$\,mag. 
Finally, \citet{Mignoli05} deals with the VLT-FORS1/FORS2 spectroscopic 
observations of $K_s$-selected galaxies from the K20 survey, 
which includes a 32.2 arcmin$^2$ field within the CDFS.

\subsection{Photometric Redshifts}

Several investigations for photometric redshifts in the CDFS region have 
been made in the past. Particularly relevant is the study by the COMBO-17
project, which has observed this region with the 
{\it Wide Field Imager} (WFI) at the ESO 2.2-m telescope with 
a 17-band filter set covering the 
$350 - 930$\,nm wavelength range. As explained by \citet{Wolf04}, 
this ``very-low resolution spectra'' allows accurate photometric redshifts to 
be obtained down to $R=23-24$\,mag. Seeing-adaptive, weighted aperture 
photometry is used, to measure the same fraction of an object in every band 
independently of seeing effects - 
for most objects this is similar to a flux measurement in an aperture of 
1.5\arcsec~diameter in 1.5\arcsec~seeing \citep{Wolf04}. 
The spectral energy distribution templates used include  star, galaxy 
(with and without extinction) and QSO types. Mean redshift and 
variance (1\,$\sigma$) as well as the classification from the best fit 
procedure were extracted from the catalog \citep{Wolf04} and 
presented here.

For optically fainter sources, eventually at higher redshifts, the 
near-infrared bands not covered by the COMBO-17 dataset assume a more 
important role. To get a photometric redshift estimate for these objects, we 
have also searched the catalog of \citet{Mobasher04}, which uses 
ground-based 
optical ($U'UBVRI$-bands) and near-infrared ($JHK_s$) data from ESO 
facilities (WFI at the 2.2-m telescope, FORS1 and ISAAC at the VLT, SOFI at 
the NTT) as well as space observations from $HST$/ACS 
($B_{435}, V_{606}, i_{775}, z_{850}$), resulting in as many as 18 independent 
photometric measurements for each source. While useful for fainter sources,
the spectral libraries used do not include QSO templates, which results in 
lack of accuracy for some sources, as can be seen in this paper. 
Also, the photometry measurements in \citet{Mobasher04} were made through 
matched 3\arcsec~diameter apertures in all bands. In the case of
some optical sources with very-close neighbors, often only identified in the 
ACS images, this can bias the photometric redshift estimate: 
we note such cases when commenting on the individual sources.

\subsection{X-ray and Radio Luminosities}

Luminosities were estimated whenever a redshift determination exists. 
In order of reliability, we have considered the spectroscopic redshift or,
in its absence, the photometric redshift from the COMBO-17 work. For faint 
sources with no other redshift information, the estimate from 
\citet{Mobasher04} was used.

The rest-frame X-ray luminosity was calculated as

\begin{equation}
L_{X} = 4\,\pi\,d_{L}^{2}\,f_{X}\,(1+z)^{\Gamma -2}~{\rm erg~s^{-1}},
\label{eqn:LX}
\end{equation} 

\noindent where $d_{L}$ is the luminosity distance (cm), $f_{X}$ is the 
X-ray flux in the 0.5-8\,keV band (${\rm erg\,cm^{-2}\,s^{-1}}$), 
and $\Gamma$ is the photon index, taken from \citet{Alexander03}. For 
sources detected only in the \citet{Giacconi02} catalogs, the  
0.5-8\,keV X-ray flux is taken as the sum of the 0.5-2\,keV and the 
2-8\,keV fluxes, where the latter is derived from the quoted 2-10\,keV flux 
assuming $\Gamma=1.4$, as explained above.

The rest-frame radio luminosity density was calculated as 

\begin{equation}
L_{\rm 1.4\,GHz} = 4\,\pi\,d_{L}^{2}\,{\rm S}_{\rm 1.4\,GHz}\,10^{-33}\,(1+z)^{\alpha -1}~{\rm W~Hz^{-1}}
\label{eqn:L1.4}
\end{equation}

\noindent where $d_{L}$ is the luminosity distance (cm), 
S$_{\rm 1.4\,GHz}$ is the 1.4\,GHz flux density (mJy), and 
$\alpha$ is the radio spectral index ($f_{\nu}\!\propto\!\nu^{-\alpha}$). 
In determining the radio luminosity density we have assumed $\alpha$ = 0.8, 
the characteristic radio spectral index of synchrotron radiation.

Of the 65 optical sources considered in Table~\ref{tab:ids}, 
34 have a spectroscopic redshift, while a further 24 
have a photometric redshift estimate. 

\subsection{Classification}

We adopt a classification based primarily on the observed X-ray
properties $L_X$ and hardness ratio ($HR$). 
$L_X$ is derived from the 0.5--8\,keV
fluxes and spectroscopic or photometric redshifts, while $HR$ is
calculated as $HR=(H-S)/(H+S)$, with $H$ and $S$ being, respectively, the
count rates in the hard (2--8\,keV) and soft (0.5--2\,keV) bands taken from
\citet{Alexander03}. For sources detected in X-rays only in the 
\citet{Giacconi02} catalogs, we assume the quoted hardness ratio 
derived from the 0.5--2\,keV and 2--10\,keV band counts, neglecting 
the (small) discrepancy arising from the different hard band used.
Our classification rests on the criteria adopted
by \citet{Szokoly04}, namely:

galaxy: $L_X<10^{42}$\,erg\,s$^{-1}$ and $HR\le -0.2$;

AGN-2: $10^{41}\le L_X<10^{44}$\,erg\,s$^{-1}$ and $HR>-0.2$;

AGN-1: $10^{42}\le L_X<10^{44}$\,erg\,s$^{-1}$ and $HR\le -0.2$;

QSO-2: $L_X \ge 10^{44}$\,erg\,s$^{-1}$ and $HR>-0.2$;

QSO-1: $L_X \ge 10^{44}$\,erg\,s$^{-1}$ and $HR\le -0.2$.

In view of the different X-ray catalog employed here, the
different bands used to calculate $L_X$ (0.5--8\,keV instead of
0.5--10\,keV) and the different K-corrections employed in the
calculation of $L_X$ we note that there are some (expected) differences 
between the X-ray classification presented in \citet{Szokoly04} and the 
one herein for some galaxies that appear in both works.

Among sources with insufficient X-ray information (or no X-ray
detection), an AGN identification was assigned whenever $L_X \ge
10^{42}$\,erg\,s$^{-1}$ (source classified as AGNX) or $L_{\rm 1.4\,GHz} \ge
10^{24.5}$\,W\,Hz$^{-1}$ (source classified as AGNR), as derived from the 
spectroscopic redshift
or the most likely photometric redshift estimate.  The adopted radio
luminosity limit is derived from the radio luminosity distribution for
star-forming galaxies and AGNs observed in the Phoenix Deep Survey
\citep{Afonso05}.

Also the optical spectroscopic information, where available, was used. 
Since most sources with a spectrum were observed by \citet{Szokoly04}, we
adopt their optical classification: BLAGN, whenever the emission lines are 
broad (full width half maximum $>$ 2000\,km\,s$^{-1}$), HEX for galaxies 
displaying narrow emission lines some of which are not found in H{\sc II} regions
(e.g., [NeV]$\lambda$3425, He{\sc II}$\lambda$1640), LEX for objects with H{\sc II} 
region-type spectrum and ABS for galaxies with an absorption line spectrum. 
Spectra from \citet{Lefevre04}, \citet{Vanzella04} and \citet{Mignoli05} were visually 
inspected and classified accordingly. Finally, diagnostic emission line 
ratios were used to separate Seyfert 2's from starforming galaxies, whenever
suitable emission lines of significant S/N were present, providing an extra 
degree of information from the optical spectra.

The final characterization of the radio sample is presented in 
Table~\ref{tab:info}, which contains the following information:

(1) source number;

(2) spectroscopic redshift and its origin;

(3) spectroscopic quality flag - given the different sources for spectroscopic 
information, we have converted quality flags from
the other works to a scale more similar to the one in \citet{Szokoly04}, 
namely: $Q_z=2.0$ denotes a reliable redshift determination, $Q_z=1.0$ 
indicates that some spectroscopic feature exists but cannot be securely
identified, and $Q_z=0.5$ flags sources for which there is only a hint of
a spectral feature;

(4) photometric redshift from the COMBO-17 work \citep{Wolf04};

(5) classification from the COMBO-17 work \citep{Wolf04};

(6) results from the photometric redshift work of \citet{Mobasher04}. The 
photometric redshift and associated 95\% 
confidence interval, the best-fit spectral galaxy type (E=1, Sbc=2, Scd=3,
IM=4, Starburst=5,6) and number of bands in which the object is
detected are presented;

(7)  $\log$ of 1.4\,GHz luminosity in W\,Hz$^{-1}$;

(8)  $\log$ of X-ray (0.5-8\,keV) luminosity in erg\,s$^{-1}$;

(9) radio-to-optical flux ratio;

(10) X-ray-to-optical flux ratio;

(11) HR, derived from the 0.5--2 and 2--8\,keV band counts. For
sources detected only in the \citet{Giacconi02} catalogs, we report the 
quoted hardness ratio, derived from the 0.5--2\,keV and 2--10\,keV band 
counts, clearly marking such cases;

(12) effective photon index $\Gamma$, as given by \citet{Alexander03};

(13) classification obtained from the X-ray characteristics ($L_X$ 
and/or $HR$)
or, if possible for sources with insufficient X-ray information, 
from the radio power;

(14) classification obtained from the optical spectrum, as in 
\citet{Szokoly04}, or after estimating any possible line-emission 
diagnostic ratios (Sey2 or SF).

\section{Comments on individual sources}

{\bf 1: ATCDFS\_J033159.86-274541.3}: Unidentified at optical 
($z_{850} > 27.7$\,mag and undetected in the stacked 
$B_{435} + V_{606} + i_{775} + z_{850}$ image) and X-ray wavelengths.

{\bf 2: ATCDFS\_J033204.81-274125.8}: The optical counterpart
matches the position of the X-ray identification. It presents a
complex optical morphology, with several peaks of brightness within a
0.5$\times$0.5\arcsec ${^2}$ region. The spectroscopic redshift
($z_{sp}=0.720$), based on a single emission line considered to be
Mg{\sc II}, is supported by the photometric redshift estimate. The high
X-ray luminosity and hard X-ray spectrum indicate a type 2 AGN. 

{\bf 3: ATCDFS\_J033205.07-274535.5}: A faint, extended optical object is
observed 2.6\arcsec~away from the radio position, at the border of the
3\,$\sigma$ radio position error region. The estimated photometric redshift 
is highly uncertain owing to the small number of optical bands where 
the object is detected. No X-ray emission is observed.

{\bf 4: ATCDFS\_J033208.51-274648.8}: A bright ($z_{850}=18.6$\,mag) 
disk galaxy at $z_{sp}=0.310$, also detected at X-rays. 
Optical spectroscopic line ratios indicate the presence of an AGN, 
confirmed by the X-ray emission properties ($L_X$ and $HR$), 
despite the low X-ray-to-optical flux ratio (source undetected in the 
soft 0.5-2\,keV X-ray band).

{\bf 5: ATCDFS\_J033208.60-274043.0}: An optically bright ($z_{850}=20.0$\,mag)
disk galaxy at $z_{ph}=0.35$.  An optically unidentified X-ray detection 
exists just outside the radio 3\,$\sigma$ error region, but it seems 
unrelated to the radio emission.


{\bf 6: ATCDFS\_J033208.67-274734.3}: An optical point-like source at
$z_{sp}=0.543$, showing indications of underlying low-level extended
optical emission. The implied high X-ray luminosity and hard X-ray
spectrum classifies this as a type-1 QSO. Broad emission lines are
detected in the optical spectrum.

{\bf 7: ATCDFS\_J033209.72-274249.0}: A spheroidal galaxy also detected in
X-rays, at $z_{sp}=0.733$. The X-ray emission is characteristic of
a type-1 AGN. Two nearby (within 20\arcsec) galaxies show similar redshifts, 
at $z_{sp}=0.728$ and $z_{sp}=0.729$. This is one of the members of the 
structure found at $z \sim 0.73$ in the CDFS \citep{Gilli03,Szokoly04,Adami05}.

{\bf 8: ATCDFS\_J033210.80-274629.2}: A compact optical source also
detected in X-rays. The redshift ($z_{sp}=1.610$) implies
high X-ray and radio luminosities characteristic of an AGN, as does
the X-ray-to-optical flux ratio.

{\bf 9: ATCDFS\_J033210.91-274415.1}: A broad emission line type-1 QSO
at $z_{sp}=1.615$ as revealed by the optical spectroscopy and X-ray
properties. Extended emission from the host galaxy is visible in the
ACS images.

{\bf 10: ATCDFS\_J033211.00-274053.6}: A compact optical source, also 
displaying X-ray emission, close (3.5\arcsec~away) to a bright spiral 
galaxy. The absence of a spectroscopic or a photometric redshift prevents 
classification of this object, although the X-ray-to-optical flux ratio is
indicative of an AGN. The observed properties of this source are
similar to those of the type-1 QSO identified for source \#9, perhaps
at a higher redshift.

{\bf 11: ATCDFS\_J033211.51-274711.5}: A galaxy with a highly irregular 
optical morphology at $z_{sp}=0.576$. No X-ray emission is detected.


{\bf 12: ATCDFS\_J033213.08-274351.0}: Despite being a relatively bright radio
source in this sample (S$_{\rm 1.4\,GHz}=1.4$\,mJy), there is no likely optical ($z_{850} > 28.2$\,mag and undetected in the stacked 
$B_{435} + V_{606} + i_{775} + z_{850}$ image) or 
X-ray counterpart. The radio-to-optical flux ratio ($>$4.86) is one of the 
largest in this sample.

{\bf 13: ATCDFS\_J033213.28-274240.3}: A complex system apparently involving a
bright ($z_{850}=19.7$\,mag) galaxy (source \#13a) with faint spiral structure
surrounded by several knots of optical emission, 
and a possible edge-on galaxy (source \#13b). This possible merger system is
identified as a single object in the public ACS GOODS catalog: 
SExtractor was used to deblend the two major components. 
The brightest source (\#13a) is an X-ray detection at a redshift of 
$z_{sp}=0.605$, with X-ray properties characteristic of a type-2 AGN. 


{\bf 14: ATCDFS\_J033214.17-274910.2}: A faint ($z_{850}=24.1$\,mag), 
extended object with a redshift of $z_{ph}=1.66$, and no X-ray detection.

{\bf 15: ATCDFS\_J033217.11-274303.9}: A spheroidal galaxy at $z_{sp}=0.569$,
with optical spectroscopy indicating a broad emission line AGN, and
the X-ray emission classifying it as a type-1 AGN.


{\bf 16: ATCDFS\_J033217.22-275222.4}: The likely optical counterpart 
is a disk galaxy at $z_{sp}=1.097$ 
with X-ray emission indicative of a type-2 AGN. The radio emission
extends to the southeast, towards a brighter source with similar spectroscopic
redshift of $z_{sp}=1.098$. The spectroscopic observations available
in a 1\arcmin$\times$1\arcmin~region centered in this source reveal
eight redshifts between 1.05 and 1.10.  This indicates the existence
of a high-redshift complex likely including also radio source \#20. 
In fact, a massive group has recently been detected in the CDFS at
this very redshift \citep{Adami05}.

{\bf 17: ATCDFS\_J033218.01-274718.4}: A galaxy at $z_{sp}=0.734$, also 
detected at X-rays. Both optical spectroscopy and X-ray properties indicate 
a star-forming galaxy, although at high X-ray and radio luminosities. The ACS
image actually resolves this source into a brighter spheroidal component 
and a much fainter companion 0.5\arcsec~away, likely revealing a merging 
event. Five nearby (within 30\arcsec) sources
have similar spectroscopic redshifts ($z_{sp}=0.73-0.74$). These are members 
of the structure known to exist in the CDFS at this redshift
\citep{Gilli03,Szokoly04,Adami05}.


{\bf 18: ATCDFS\_J033218.03-275056.2}: A faint optical source with a 
complex morphology lies inside the radio position error region. The
photometric redshift ($z_{ph}=1.04$), essentially from ground-based 
photometry, is likely to be biased by a slightly brighter source 
0.4\arcsec~to the south, whose relation to the assumed optical ID is unclear.
A nearby bright disk galaxy and the X-ray detection east of the 
radio coordinates are too distant to be considered related to the radio 
emission.


{\bf 19: ATCDFS\_J033218.83-275410.3}: No optical identification exists 
within the 3\,$\sigma$ radio position error region. Instead, this radio 
source is thought to form, with source \#21, the 
double radio lobe structure related to a cD-type galaxy 
10.2\arcsec~away (bright galaxy indicated by the horizontal lines), 
at a photometric redshift of $z_{ph}=0.96$. A weak X-ray detection from the 
\citet{Giacconi02} catalog is associated to the optical galaxy, 
indicating the presence of a type-2 AGN.
From the available redshift surveys in this field, there are four nearby 
galaxies at similar redshifts ($z_{sp}=0.97-1.04$), which indicates the 
presence of a group. It is unclear the association with another
weak X-ray detection, 3\arcsec~south from the radio position 
(see Figure~\ref{fig:pstamps}), again detected only in the 
\citet{Giacconi02} catalog (identified as source 527 in that catalog). 


{\bf 20: ATCDFS\_J033219.41-275216.5}: An irregular galaxy showing 
indications for a tidal tale towards the southeast, which suggests
a merger event. The photometric redshift $z_{ph}=1.05$ is comparable to the 
spectroscopic redshift of 1.096 for the slightly fainter galaxy 
3\arcsec~to the southwest, a possible merger companion. The redshifts 
suggest membership of the structure discussed for source \#16.  

{\bf 21: ATCDFS\_J033219.57-275403.2}: This source is thought to be
the second radio lobe associated to a cD-type galaxy 
5.5\arcsec~away (the bright galaxy indicated by the horizontal lines). 
See the discussion for source \#19, the other proposed radio lobe.

{\bf 22: ATCDFS\_J033219.82-274121.2}: A galaxy at $z_{sp}=0.229$,
also detected at X-rays. The X-ray emission does not provide a definite 
classification of this source, although the optical line ratio points to 
a Seyfert 2 type galaxy. The X-ray-to-optical flux ratio, however, is  
more indicative of star-formation. 

{\bf 23: ATCDFS\_J033221.24-274436.1}: The optical identification seems to be 
part of a merging pair, the second galaxy just outside the radio 
position error region. The optical spectrum, revealing a redshift of $z_{sp}=0.524$, likely  
includes light from both sources, 0.9\arcsec~apart. This is also a weak X-ray
detection in the 0.5-1\,keV band \citep[source S12 of][]{Alexander03}.
The low X-ray-to-optical flux ratio is indicative of star-formation.

{\bf 24: ATCDFS\_J033221.92-274243.8}: A weak extended optical source. The
photometric redshift of $z_{ph}=1.54$ may be biased by a brighter
source 1\arcsec~to the north, whose relation to the assumed optical ID is
unclear.


{\bf 25: ATCDFS\_J033222.36-274807.3}:  Unidentified at optical 
($z_{850} > 28.4$\,mag and undetected in the stacked 
$B_{435} + V_{606} + i_{775} + z_{850}$ image) and X-ray wavelengths.
The nearby X-ray detections seem unrelated to the radio emission.

{\bf 26: ATCDFS\_J033222.64-274424.5}: A spiral galaxy at $z_{sp}=0.737$,
displaying weak X-ray emission detected only in the 0.5-8\,keV band
\citep[source 148 of][]{Alexander03}. The optical spectroscopy, although 
of low S/N, is indicative of star-formation, as is the 
low X-ray-to-optical flux ratio. This is another likely member of the 
structure detected at this redshift in the CDFS 
\citep{Gilli03,Szokoly04,Adami05}.


{\bf 27: ATCDFS\_J033224.59-275441.6}: A disk galaxy at $z_{sp}=0.123$, 
with no X-ray emission. The available optical spectroscopy does not provide 
enough lines to apply line ratio diagnostics. 

{\bf 28: ATCDFS\_J033225.15-275452.2}: A ellipsoidal galaxy with a 
spectroscopic redshift of $z_{sp}=1.090$, also detected at X-rays. 
The X-ray emission classifies it as a type-2 AGN,  while the optical 
spectroscopy reveals an absorption-line spectrum. This is a likely 
member of the structure recently found at $z \sim 1.10$ in the CDFS 
\citep{Adami05}.


{\bf 29: ATCDFS\_J033226.98-274106.5}: This is the brightest radio source in 
the ACS area, with S$_{\rm 1.4\,GHz}=20$\,mJy. The optical identification is
a point-like source which shows the brightest X-ray flux in this sample. At
$z_{sp}=0.734$, this is a type-1 QSO, with optical spectroscopy revealing a 
broad-line AGN. At least three other galaxies lie at the
same redshift in the surrounding 1\arcmin$\times$1\arcmin~region, again
tracing the structure detected at this redshift in the CDFS 
\citep{Gilli03,Szokoly04,Adami05}.

{\bf 30: ATCDFS\_J033227.64-275040.5}: A galaxy at $z_{sp}=1.097$,
also detected in X-rays. The optical image reveals a very irregular,
knotty morphology, possibly indicative of a merging event and 
related star-formation. 
The X-ray emission does not elucidate on the processes behind its luminosity, 
although the X-ray-to-optical flux ratio is typical of star-forming galaxies. 
The optical spectroscopy presents some lines which are more common on AGNs 
(e.g., [NeV]$\lambda 3425$). However, this line may also be generated in 
shocks in interacting galaxies. This radio source is another likely member 
of the massive group recently detected in the CDFS at $z \sim 1.10$ 
\citep{Adami05}.

{\bf 31: ATCDFS\_J033227.98-274641.5}: A bright ($z_{850}=19.4$\,mag) disk 
galaxy at $z_{sp}=0.247$. This source was 
observed spectroscopically by \citet{Croom01}, revealing a spectra dominated 
by absorption features, although it presents emission in H$\alpha$ and 
[N{\sc II}]. A fainter optical source, which could be a companion galaxy, 
is seen 1.5\arcsec~to the north

{\bf 32: ATCDFS\_J033228.31-273842.9}: A galaxy at $z_{ph}=0.39$ presenting
irregular structure with a nearby (1.7\arcsec~away), possibly related,
companion.

{\bf 33: ATCDFS\_J033228.71-274402.3}: Unidentified at optical 
($z_{850} > 28.1$\,mag and undetected in the stacked 
$B_{435} + V_{606} + i_{775} + z_{850}$ image) and X-ray wavelengths.

{\bf 34: ATCDFS\_J033228.78-274620.8}: An ellipsoidal galaxy at a redshift of
$z_{sp}=0.738$, also detected in X-rays. Several nearby fainter optical
sources exist and low-level optical emission bridges to at least one
neighbor, which suggests interaction. The galaxy is classified as a type-1
AGN from its X-ray properties. Again, this is a likely member of the 
$z \sim 0.73$ structure that exists in the CDFS 
\citep{Gilli03,Szokoly04,Adami05}.

{\bf 35: ATCDFS\_J033229.59-274332.5}: Unidentified at optical 
($z_{850} > 28.1$\,mag and undetected in the stacked 
$B_{435} + V_{606} + i_{775} + z_{850}$ image) and X-ray wavelengths.

{\bf 36: ATCDFS\_J033229.83-274423.7}: A flocculent galaxy, 
with X-ray emission coincident with its nuclear region. At
$z_{sp}=0.076$ the galaxy is part of a larger structure - 
low-level optical emission connects this galaxy to the optical identification
of source \#38, indicating an interacting pair \citep[as previously 
noted by][]{Giacconi01}, and a smaller galaxy at the same redshift is 
located roughly 20\arcsec~west. The X-ray characteristics
indicate a star-forming galaxy.

{\bf 37: ATCDFS\_J033229.89-274520.0}: A spiral/flocculent galaxy at
$z_{sp}=0.953$. X-ray emission detected towards the west side of the
galaxy could be related to an area of brighter optical emission in the
western spiral arm. The X-ray and radio characteristics do not reveal
AGN activity. Two other galaxies in the immediate
vicinity (within 9\arcsec) lie at similar redshifts, indicating the
presence of a larger structure.

{\bf 38: ATCDFS\_J033229.96-274404.8}: Another flocculent galaxy at
$z_{sp}=0.076$, connected by low-level optical emission to source 
\#36. X-ray emission is detected north of the nuclear
region, with characteristics indicative of star formation. 

{\bf 39: ATCDFS\_J033229.98-275258.7}: An extended faint optical source 
($z_{850}=26.19$), is identified as the possible optical counterpart. 
However, this is a complex region, both in the optical and in the radio: 
a couple of faint optical sources appear near the radio position in the 
combined $B_{435} + V_{606} + i_{775} + z_{850}$ image, and the radio 
emission partly overlaps source \#40. In fact, 
an X-ray source associated with an optically complex multiple system is
observed halfway between
sources \#39 and \#40: a possible interpretation would be that these 
two radio sources are somehow associated with the X-ray source 
\citep[source S22 of][]{Alexander03}, although the X-ray characteristics 
and optical spectroscopy do not indicate any AGN activity.

{\bf 40: ATCDFS\_J033230.20-275312.7}: The most likely optical counterpart 
is a very faint ($z_{850}=26.8$\,mag) galaxy revealed in the combined 
$B_{435} + V_{606} + i_{775} + z_{850}$ image as a pair of sources.
It should be noted that this 
is a complex region in both the optical and the radio, with some 
overlapping between the radio emission  of sources \#39 and \#40, which 
could even be related; see the discussion for source \#39.

{\bf 41: ATCDFS\_J033231.44-274621.5}: The radio source has two candidate
optical IDs with similar values of $\mathcal{R}$. The most likely 
(source \#41a) is an X-ray source with a spectroscopic 
redshift $z_{sp}=2.223$, which agrees with the photometric redshift estimate. 
This source has a high X-ray 
luminosity indicating an AGN \citep[see also][]{Szokoly04}, although
\citet{Daddi04} argues that it may be a vigorous star-forming
galaxy. Recently, \citet{vanDokkum05} performed Gemini near-infrared 
spectroscopy of this source, and found a surprising lack of AGN spectral 
features. Instead, the observed Seyfert-like line ratios could be 
explained by shock ionization due to a strong galactic wind in a starburst 
galaxy.
A second possible identification (source \#41b) has no detectable X-ray 
emission and a photometric redshift estimate of $z_{ph}=0.84$. The optical 
image shows an extended galaxy, hinting at the existence of tidal tail 
disruption.

{\bf 42: ATCDFS\_J033233.44-275228.0}: Unidentified at optical 
($z_{850} > 28.1$\,mag and undetected in the stacked 
$B_{435} + V_{606} + i_{775} + z_{850}$ image) wavelengths. A weak 
X-ray detection exists in the \citet{Giacconi02} catalog.

{\bf 43: ATCDFS\_J033234.97-275456.1}: A faint object is just perceptible
in the $z_{850}$ image, although it is not present in the public ACS GOODS
catalogs. In the stacked $B_{435} + V_{606} + i_{775} + z_{850}$ image,
two faint objects are visible. Inspection of the individual 
optical bands reveal one of these sources appearing only in the bluer bands 
($B_{435}$ and $V_{606}$), while the other appears mainly in the $V_{606}$
and $i_{775}$ band, being just hinted in the redder band at
$z_{850}\sim 27.5$ (value from aperture photometry). 

{\bf 44: ATCDFS\_J033235.07-275532.8}: A disk galaxy at $z_{sp}=0.038$ with 
X-ray emission originating from the disk. As discussed by \citet{Szokoly04},
the X-ray luminosity, hardness ratio and the location of the X-ray
emission indicate a ultraluminous X-ray source (ULX) hosted by a normal
galaxy. The low radio luminosity is consistent with this interpretation.

{\bf 45: ATCDFS\_J033235.46-275452.8}: A faint optical source 
($z_{850}=26.8$\,mag) is identified as the possible counterpart,
although the stacked $B_{435} + V_{606} + i_{775} + z_{850}$ image 
hints at the presence of fainter sources in the region. 
The optical counterpart candidate displays a faint elongated morphology. 
No X-ray emission is present in this region.

{\bf 46: ATCDFS\_J033235.71-274916.0}: A compact optical source with a
close neighbor only visible in the ACS images (0.3\arcsec~away) is the probable identification, at $z_{sp}=2.578$. X-ray emission is also present. The radio and X-ray properties classify this as a type-2 AGN.

{\bf 47: ATCDFS\_J033237.22-275129.7}: A pair of interacting galaxies at
a photometric redshift, which includes light from both sources, of 
$z_{ph}=0.53$. The western galaxy is a face-on spiral (source \#47a) while 
the eastern source is disk galaxy (source \#47b). 
No X-ray emission is detected.

{\bf 48: ATCDFS\_J033237.75-275000.7}: A complex optical system at
$z_{ph}=1.36$, with this photometric redshift estimate including light from the
entire region. Two optical sources separated by around 1\arcsec~are
identified as likely counterparts. Careful inspection of the stacked 
$B_{435} + V_{606} + i_{775} + z_{850}$ image
reveals four patches of optical emission surrounded by elongated structures of 
low surface brightness, perhaps indicating a merger event. 
No X-ray emission is detected.

{\bf 49: ATCDFS\_J033237.75-275211.9}: A compact optical source 
also detected in the X-rays with $z_{sp}=1.603$. The X-ray properties 
(luminosity and hardness ratio) indicate a type-1 QSO.

{\bf 50: ATCDFS\_J033238.75-274632.5}: The possible optical identification for
this weak radio source is a bright ($z_{850}=21.533$\,mag) galaxy at 
$z_{ph}=0.57$. However, we note that this optical identification is of low
reliability ($\mathcal{R}=0.23$), and the slightly extended east-west radio 
contours rise the possibility of complex radio emission originating in more 
than one of the galaxies seen in this region.

{\bf 51: ATCDFS\_J033239.46-275300.9}: Two ellipsoidal galaxies 
separated by 1.2\arcsec~in
a likely merging event (as indicated by the low surface brightness
extended emission). The photometric redshift is $z_{ph}=0.65$, identical for 
both sources, 
but resulting from separate analysis in the \citet{Wolf04} work.

{\bf 52: ATCDFS\_J033239.64-274851.8}: Two unrelated optical sources are
possible counterparts. The most likely counterpart (source \#52a) is an 
X-ray source with $z_{sp}=3.064$, classified as a type-2 AGN from its 
X-ray properties. The other possible counterpart (source \#52b) has no X-ray 
emission and a photometric redshift of $z_{ph}=1.40$.

{\bf 53: ATCDFS\_J033240.82-275545.8}: A faint optical source 
($z_{850}=25.18$\,mag) with X-ray emission. 
A spectrum with a ``hint of some spectral feature'' was
obtained for this source by \citet{Szokoly04}, suggesting a redshift of
$z_{sp}=0.625$. The photometric estimate results in a higher value of
$z_{ph}=1.81$. Considering even the smallest (spectroscopic) value, the
X-ray characteristics indicate a type-2 AGN consistent with the
X-ray-to-optical flux ratio.

{\bf 54: ATCDFS\_J033243.12-275514.2}: A pair of galaxies separated by 
0.6\arcsec~with disturbed morphology associated with traces of low 
surface brightness tails, indicating a merging system. The photometric
redshift, based on the light from both sources, is $z_{ph}=1.63$. No
X-ray emission is detected.

{\bf 55: ATCDFS\_J033244.16-275142.4}: A spectroscopically confirmed pair of
galaxies at $z_{sp}=0.279$. The western companion (source \#55a) is
orientated edge-on and has clear signs of dust obscuration, 
displaying an optical spectrum dominated by absorption features.
The eastern galaxy (source \#55b) displays an emission-line optical 
spectrum with 
inconclusive line ratios. This latter galaxy is an X-ray source, with X-ray 
characteristics indicative of star-formation. However, the optical 
spectroscopy reveals line ratios suggesting a Seyfert 2 type galaxy.

{\bf 56: ATCDFS\_J033244.93-274726.2}: A spiral galaxy at $z_{sp}=0.214$, with
no X-ray emission and narrow emission-line optical spectrum.

{\bf 57: ATCDFS\_J033245.03-275438.8}: A face-on spiral galaxy with
$z_{sp}=0.458$, showing knots of star formation. Faint X-ray emission
is detected, possibly off-centered towards the north arm of the
galaxy. The X-ray characteristics are consistent with star
formation.

{\bf 58: ATCDFS\_J033246.03-275318.2}: No reliable optical counterpart is
detected for this radio source, even in the stacked 
$B_{435} + V_{606} + i_{775} + z_{850}$ image. No X-ray emission is observed 
either. One should note that this is a complex region in the radio, since the 
radio emission overlaps with that of source \#59.

{\bf 59: ATCDFS\_J033246.33-275328.6}: The likely optical counterpart 
exhibits a complex morphology and an estimated photometric redshift
of $z_{ph}=0.833$. No X-ray emission is detected.

{\bf 60: ATCDFS\_J033246.78-275120.1}: Two galaxies without X-ray emission 
are possible optical counterparts, with similar values for $\mathcal{R}$.
Photometry indicates $z_{ph}=1.06$ for the southern object (source \#60a), 
while the northern galaxy (source \#60b) has no reliable 
redshift determination.

{\bf 61: ATCDFS\_J033248.02-275414.7}: A faint optical source 
($z_{850}=26.57$\,mag) lies close to the radio position and is the possible
optical counterpart. The faint magnitude prevents any photometric redshift 
estimate.

{\bf 62: ATCDFS\_J033248.62-274934.9}: An extended object at a spectroscopic 
redshift of $z_{sp}=1.117$ which, despite based on a low S/N spectrum, 
is supported by the photometric redshift estimates. A second optical source 
lies less than 1\arcsec~to the southeast, just outside the 3\,$\sigma$ radio 
position error region. Traces of low surface brightness emission connect both
sources, which suggests an interacting system. This is also a weak X-ray 
detection in the \citet{Giacconi02} catalog. 

{\bf 63: ATCDFS\_J033300.91-275049.4}: A bright 
($z_{850}=21.67$\,mag) optical source is identified as the possible 
optical counterpart, although at a relativelly low value of $\mathcal{R}$.
No X-ray emission is detected from this region.

{\bf 64: ATCDFS\_J033302.99-275147.3}: An X-ray counterpart exists with a weak
optical source ($z_{850}=25.13$\,mag) lying just outside the 
position error box of the X-ray detection. 
The photometric redshift of the optical source is $z_{ph}=0.45$, as estimated
by \citet{Mobasher04}. However, for this same optical source, \citet{Zheng04} 
finds $z_{ph}=3.69$, using other photometric redshift codes. If the optical 
and X-ray emission are from the same source, the X-ray characteristics 
indicate, at the lowest redshift mentioned above, a type-1 AGN. 
If the higher redshift is the correct one, then this will be classified as a 
type-1 QSO.

\section{Discussion}

There are 64 sources in the radio catalog. Likely ACS counterparts have 
been identified for 50 of these (78\% - where we include unique optical 
counterparts with $\mathcal{R} \geq 0.5$, likely merger systems and 
sources \#19 and \#21 as the radio lobes of the same radio galaxy). 
Optical identifications with lower reliability ($\mathcal{R} < 0.5$)  
exist for 7 radio sources, including three cases where seemingly independent
optical identifications are equally reliable at the $\mathcal{R} \sim 0.5$ 
level (sources \#41, \#52 and \# 60). 
For 7 radio sources (11\%) no optical identification was possible.

The redshifts for the optically identified radio sources range from below 0.1
to above 3, although most are found up to $z \sim 1.1$. This is in agreement 
with previous work on the Hubble Deep Field North (HDF-N) and the 
Small Selected Area (SSA) 13 fields, which reach similar depths for
the spectroscopic observations \citep{Chapman03}. Several radio sources 
are identified at $z \sim 0.73$ and $z \sim 1.10$. 
These are likely members of the structures already found in the 
CDFS region \citep{Gilli03,Szokoly04,Adami05}.

A total
of 34 radio sources (53\%) have an X-ray counterpart - in all but three cases,
the X-ray identifications are unambiguously associated with reliable optical 
counterparts of the radio sources. For sources \#41 and \#52, where two optical
sources appear as possible counterparts, the X-ray detection suggests 
which optical counterpart is the real one. Source \#42 has no optical counterpart despite the X-ray detection.

The 7 radio sources without ACS counterparts (hereinafter ``optically faint 
radio sources'') are of particular interest.  In a pioneering study of 
optically faint sub-milliJansky radio sources, \citet{Richards99}
reported that approximately 60\% 
of the radio sources detected in very deep 1.4~GHz surveys of the 
Hubble Deep Field (North) and the Small Selected Area 13 fields, 
have relatively bright ($I \sim 22$\,mag) disk galaxy counterparts, 20\%
have low-luminosity AGN counterparts, and the remaining 20\% 
have no counterpart brighter than $I=25$.  Deep $K$-band imaging of this
population revealed that several are very red objects, with $I-K > 4$ 
\citep{Richards99}, suggesting that these objects form an inhomogeneous
population, including high-redshift dusty starbursts, extreme redshift
radio-loud AGNs, moderate redshift AGNs with optical hosts having a relatively
low optical power, or one-sided radio jets.  With the deep ACS imaging
reported here, we find 16 sources (25\%) with $z_{850} \geq 25$\,mag, with 
the fraction of optically unidentified ($z_{850} \gtrsim 28$\,mag) radio 
sources being 11\% (7 sources). However, the radio flux density of 3 out of 
these 7 optically unidentified sources is brighter 
than 1\,mJy, significantly brighter than the sources studied by 
\citet{Richards99}. These sources are either extremely reddened, or are 
extremely "radio loud". They invite follow-up infrared imaging to 
elucidate their character.


Altogether, there are 23 radio sources (counting sources \#19 and \#21 as 
a single one) for which a classification is
possible based on the X-ray emission alone. Of these, 15 (65\%) have an AGN 
designation, 4 (17\%) are QSOs, and 4 are classified as
normal galaxies. Thus, based on the properties of X-ray counterparts, 
19 (30\%) of the radio sources have evidence of an AGN. 
Considering only the sub-mJy 
radio sources in the sample, then 15 out of 55 sources (27\%) show signs of 
AGN activity based on their detected X-ray emission alone. 
The proportion of faint radio sources associated with AGN X-ray sources is 
thus significantly higher that the proportion (20\%) of radio sources 
with evidence of AGN activity reported by \citet{Richards99} on the 
basis of classification by radio spectral index and optical morphology.  

It is claimed that very deep radio images will provide a population of
radio-selected objects that are of great astrophysical interest, since they
will be selected without the intervention of dust obscuration.  The present
survey lends weight to this claim.  The optical and X-ray counterparts of the
faint radio sources, as well as the blank fields, invite further study at
infrared wavelengths since many will be galaxies exhibiting evidence of
violent star formation and AGN activity at intermediate and high redshifts.

\section{Conclusions}

We present the cross-correlation between a deep radio survey of the 
GOODS-S ACS field and the available optical and X-ray data. 
Out of the 64 radio sources in this region, 58 have an optical and/or X-ray 
identification. Spectroscopic or photometric redshifts are available for
the majority of these sources, which allows a classification to be 
established.

The superb ACS imaging quality reveals the morphological characteristics 
of the radio sample which can now be used, in particular, to analyze in more
detail the morphology of radio-selected star-forming galaxies.

In spite of the very deep optical observations available, seven radio 
sources remain optically unidentified, all but one revealing no indications 
of X-ray emission. Infrared observations, in particular those being done 
by the Spitzer Space Telescope, will probably reveal some of these sources, 
likely to be very dusty star-forming galaxies or very high redshift AGN. 
Further radio observations will also be needed to elucidate on the physical
mechanisms present in these sources.

\acknowledgments

JA gratefully acknowledges the support from the Science and Technology
Foundation (FCT, Portugal) through the research grant 
POCTI/CTE-AST/58027/2004. We thank the anonymous referee for 
insightful and constructive comments that have improved this paper.

\clearpage

\begin{deluxetable}{rrrcrrrcrrr}
\tabletypesize{\tiny}
\rotate
\tablecolumns{11}
\tablewidth{0pc}
\tablecaption{Identification of faint radio sources \label{tab:ids}}
\tablehead{
\colhead{No} & \colhead{Radio source} &
\colhead{${\rm S}_{\rm 1.4\,GHz}$} &
  \colhead{ACS} & \colhead{$\mathcal{R}$} & \colhead{dist} & \colhead{$z_{850}$} &
  \colhead{{CXO}} & \colhead{dist} &
  \colhead{$f_{0.5-2}$} & \colhead{$f_{2-8}$} \\
\colhead{} & \colhead{} &
\colhead{(mJy)} & 
  \colhead{} & \colhead{} & \colhead{\arcsec} &              \colhead{} & 
  \colhead{} & \colhead{\arcsec} &
  \multicolumn{2}{c}{($10^{-18}$\,W\,m$^{-2}$)} 
}
\startdata
 1  & ATCDFS\_J033159.86-274541.3 &     0.137 $\pm$  0.042 &     \nodata         & \nodata  & \nodata  & $>$27.7  &     \nodata         & \nodata  &    \nodata  &    \nodata \\
 2  & ATCDFS\_J033204.81-274125.8 &     0.095 $\pm$  0.038 & J033204.84-274127.4 & 0.86     &  1.6     &  23.545  & J033204.89-274127.6 &  2.0     & $<$ 0.23    &     5.66   \\
 3  & ATCDFS\_J033205.07-274535.5 &     0.087 $\pm$  0.042 & J033205.21-274537.3 & 0.38     &  2.6     &  25.300  &     \nodata         & \nodata  &    \nodata  &    \nodata \\
 4  & ATCDFS\_J033208.51-274648.8 &     0.193 $\pm$  0.040 & J033208.53-274648.3 & 1.00     &  0.6     &  18.565  & J033208.50-274648.6 &  0.3     & $<$ 0.09    &     1.14   \\
 5  & ATCDFS\_J033208.60-274043.0 &     0.077 $\pm$  0.038 & J033208.67-274042.9 & 0.98     &  0.8     &  20.048  &     \nodata         & \nodata  &    \nodata  &    \nodata \\
 6  & ATCDFS\_J033208.67-274734.3 &     1.958 $\pm$  0.046 & J033208.66-274734.4 & 1.00     &  0.2     &  18.483  & J033208.66-274734.4 &  0.2     &    44.61    &    67.71   \\
 7  & ATCDFS\_J033209.72-274249.0 &     0.229 $\pm$  0.039 & J033209.71-274248.1 & 0.99     &  0.9     &  20.479  & J033209.69-274248.4 &  0.7     &     0.42    & $<$ 0.62   \\
 8  & ATCDFS\_J033210.80-274629.2 &     0.157 $\pm$  0.042 & J033210.79-274627.8 & 0.98     &  1.5     &  22.911  & J033210.80-274627.6 &  1.7     &     0.09    & $<$ 0.38   \\
 9  & ATCDFS\_J033210.91-274415.1 &     3.057 $\pm$  0.052 & J033210.91-274414.9 & 1.00     &  0.2     &  22.371  & J033210.91-274415.1 &  0.1     &     6.14    &     9.81   \\
10  & ATCDFS\_J033211.00-274053.6 &     0.286 $\pm$  0.038 & J033210.99-274053.7 & 0.99     &  0.2     &  23.577  & J033211.00-274053.7 &  0.1     &     1.88    &     3.94   \\
11  & ATCDFS\_J033211.51-274711.5 &     0.137 $\pm$  0.040 & J033211.50-274713.1 & 0.99     &  1.5     &  21.086  &     \nodata         & \nodata  &    \nodata  &    \nodata \\
12  & ATCDFS\_J033213.08-274351.0 &     1.366 $\pm$  0.038 &     \nodata         & \nodata  & \nodata  & $>$28.2  &     \nodata         & \nodata  &    \nodata  &    \nodata \\
13a & ATCDFS\_J033213.28-274240.3 &     0.127 $\pm$  0.048 & J033213.23-274241.0\tablenotemark{a} & 0.91     &  0.9     &  19.698  & J033213.24-274240.9 &  0.8     &     2.48    &    18.73   \\
13b &                             &                        & J033213.37-274239.9\tablenotemark{a} & \nodata  &  1.3     &  21.919  &     \nodata         & \nodata  &    \nodata  &    \nodata \\
14  & ATCDFS\_J033214.17-274910.2 &     0.141 $\pm$  0.034 & J033214.13-274910.1 & 0.97     &  0.6     &  24.105  &     \nodata         & \nodata  &    \nodata  &    \nodata \\
15  & ATCDFS\_J033217.11-274303.9 &     0.079 $\pm$  0.040 & J033217.14-274303.3 & 0.98     &  0.7     &  20.566  & J033217.14-274303.3 &  0.7     &     4.50    &     6.39   \\
16  & ATCDFS\_J033217.22-275222.4 &     0.063 $\pm$  0.032 & J033217.17-275220.8 & 0.70     &  1.8     &  21.766  & J033217.18-275220.9 &  1.7     &     0.55    &    20.15   \\
17  & ATCDFS\_J033218.01-274718.4 &     0.404 $\pm$  0.034 & J033218.01-274718.5 & 1.00     &  0.1     &  19.405  & J033218.07-274718.2 &  0.7     &     0.28    & $<$ 0.32   \\
18  & ATCDFS\_J033218.03-275056.2 &     0.203 $\pm$  0.048 & J033218.17-275056.6 & 0.43     &  1.8     &  26.263  &     \nodata         & \nodata  &    \nodata  &    \nodata \\
19  & ATCDFS\_J033218.62-275411.4 &     4.779 $\pm$  0.071 & J033219.29-275406.1 & \nodata  & 10.2     &  20.279  &       (G249)\tablenotemark{c} & 12.4     &     0.15    &     0.61   \\
20  & ATCDFS\_J033219.41-275216.5 &     0.201 $\pm$  0.032 & J033219.52-275217.7 & 0.55     &  1.8     &  21.871  &     \nodata         & \nodata  &    \nodata  &    \nodata \\
21  & ATCDFS\_J033219.62-275402.9 &     5.861 $\pm$  0.083 & J033219.29-275406.1 & \nodata  &  5.5     &  20.279  &       (G249)\tablenotemark{c} &  4.0     &     0.15    &     0.61   \\
22  & ATCDFS\_J033219.82-274121.2 &     0.228 $\pm$  0.085 & J033219.81-274122.7 & 0.98     &  1.5     &  18.736  & J033219.81-274123.1 &  1.9     &     0.22    & $<$ 0.94   \\
23  & ATCDFS\_J033221.24-274436.1 &     0.178 $\pm$  0.034 & J033221.28-274435.6 & 0.97     &  0.7     &  19.625  & J033221.24-274435.9 &  0.2     & $<$ 0.08    & $<$ 0.31   \\
24  & ATCDFS\_J033221.92-274243.8 &     0.089 $\pm$  0.040 & J033222.01-274243.3 & 0.80     &  1.2     &  24.099  &     \nodata         & \nodata  &    \nodata  &    \nodata \\
25  & ATCDFS\_J033222.36-274807.3 &     0.144 $\pm$  0.042 &     \nodata         & \nodata  & \nodata  & $>$28.4  &     \nodata         & \nodata  &    \nodata  &    \nodata \\
26  & ATCDFS\_J033222.64-274424.5 &     0.097 $\pm$  0.034 & J033222.58-274425.8 & 0.97     &  1.5     &  20.279  & J033222.63-274426.0 &  1.5     & $<$ 0.07    & $<$ 0.43   \\
27  & ATCDFS\_J033224.59-275441.6 &     0.086 $\pm$  0.038 & J033224.53-275443.0 & 0.96     &  1.6     &  18.907  &     \nodata         & \nodata  &    \nodata  &    \nodata \\
28  & ATCDFS\_J033225.15-275452.2 &     0.079 $\pm$  0.038 & J033225.16-275450.1 & 0.97     &  2.1     &  21.208  & J033225.17-275449.6 &  2.7     & $<$ 0.08    &     3.91   \\
29  & ATCDFS\_J033226.98-274106.5 &    20.011 $\pm$  0.208 & J033227.01-274105.0 & 0.73     &  1.6     &  19.004  & J033227.00-274105.1 &  1.5     &    49.41    &    68.58   \\
30  & ATCDFS\_J033227.64-275040.5 &     0.076 $\pm$  0.032 & J033227.72-275040.8 & 0.92     &  1.0     &  21.420  & J033227.67-275040.7 &  0.3     &     0.03    & $<$ 0.27   \\
31  & ATCDFS\_J033227.98-274641.5 &     0.094 $\pm$  0.030 & J033227.99-274639.2 & 0.96     &  2.3     &  19.364  &     \nodata         & \nodata  &    \nodata  &    \nodata \\
32  & ATCDFS\_J033228.31-273842.9 &     0.142 $\pm$  0.060 & J033228.35-273841.7 & 0.99     &  1.4     &  19.246  &     \nodata         & \nodata  &    \nodata  &    \nodata \\
33  & ATCDFS\_J033228.71-274402.3 &     4.061 $\pm$  0.083 &     \nodata         & \nodata  & \nodata  & $>$28.1  &     \nodata         & \nodata  &    \nodata  &    \nodata \\
34  & ATCDFS\_J033228.78-274620.8 &     0.237 $\pm$  0.036 & J033228.74-274620.4 & 0.98     &  0.7     &  21.017  & J033228.73-274620.2 &  0.9     &     0.25    &     0.20   \\
35  & ATCDFS\_J033229.59-274332.5 &     1.429 $\pm$  0.076 &     \nodata         & \nodata  & \nodata  & $>$28.1  &     \nodata         & \nodata  &    \nodata  &    \nodata \\
36  & ATCDFS\_J033229.83-274423.7 &     1.088 $\pm$  0.044 & J033229.88-274424.4 & 0.97     &  0.8     &  16.451  & J033229.88-274425.0 &  1.3     &     0.81    & $<$ 0.34   \\
37  & ATCDFS\_J033229.89-274520.0 &     0.115 $\pm$  0.054 & J033229.85-274520.5 & 0.91     &  0.7     &  21.009  & J033229.75-274520.3 &  1.9     & $<$ 0.07    &     0.13   \\
38  & ATCDFS\_J033229.96-274404.8 &     0.449 $\pm$  0.036 & J033229.99-274404.8 & 1.00     &  0.3     &  16.840  & J033230.01-274404.0 &  1.0     &     0.57    &     0.71   \\
39  & ATCDFS\_J033229.98-275258.7 &     0.082 $\pm$  0.025 & J033230.10-275300.1 & 0.40     &  2.0     &  26.190  &     \nodata         & \nodata  &    \nodata  &    \nodata \\
40  & ATCDFS\_J033230.20-275312.7 &     0.063 $\pm$  0.024 & J033230.23-275312.4 & 0.62     &  0.5     &  26.759  &     \nodata         & \nodata  &    \nodata  &    \nodata \\
41a & ATCDFS\_J033231.44-274621.5 &     0.063 $\pm$  0.036 & J033231.46-274623.2 & 0.49     &  1.6     &  22.877  & J033231.47-274623.0 &  1.4     &     0.06    & $<$ 0.33   \\
41b &                             &                        & J033231.40-274621.4 & 0.48     &  0.6     &  23.046  &     \nodata         & \nodata  &    \nodata  &    \nodata \\
42  & ATCDFS\_J033233.44-275228.0 &     0.097 $\pm$  0.032 &     \nodata         & \nodata  & \nodata  & $>$28.1  &       (G632)\tablenotemark{c} &  1.5     & $<$ 0.06    &     0.46   \\
43  & ATCDFS\_J033234.97-275456.1 &     0.127 $\pm$  0.030 & J033234.96-275455.8\tablenotemark{b} & 0.98     &  0.4     & 27.482   &     \nodata         & \nodata  &    \nodata  &    \nodata \\
44  & ATCDFS\_J033235.07-275532.8 &     0.310 $\pm$  0.048 & J033235.08-275533.0 & 1.00     &  0.2     &  15.698  & J033234.73-275533.8 &  4.6     & $<$ 0.13    &     1.66   \\
45  & ATCDFS\_J033235.46-275452.8 &     0.078 $\pm$  0.028 & J033235.51-275449.9 & 0.56     &  3.0     &  26.166  &     \nodata         & \nodata  &    \nodata  &    \nodata \\
46  & ATCDFS\_J033235.71-274916.0 &     0.072 $\pm$  0.030 & J033235.71-274916.0 & 0.97     &  0.1     &  24.660  & J033235.72-274916.0 &  0.1     &     0.05    &     0.70   \\
47a & ATCDFS\_J033237.22-275129.7 &     0.077 $\pm$  0.032 & J033237.17-275127.9 & 0.90     &  1.9     &  21.352  &     \nodata         & \nodata  &    \nodata  &    \nodata \\
47b &                             &                        & J033237.34-275127.4 & \nodata  &  2.8     &  21.028  &     \nodata         & \nodata  &    \nodata  &    \nodata \\
48a & ATCDFS\_J033237.75-275000.7 &     0.181 $\pm$  0.038 & J033237.74-275000.4 & 0.51     &  0.4     &  23.353  &     \nodata         & \nodata  &    \nodata  &    \nodata \\
48b &                             &                        & J033237.76-275001.4 & 0.48     &  0.6     &  23.310  &     \nodata         & \nodata  &    \nodata  &    \nodata \\
49  & ATCDFS\_J033237.75-275211.9 &     0.091 $\pm$  0.032 & J033237.76-275212.3 & 0.98     &  0.3     &  23.540  & J033237.77-275212.4 &  0.5     &     5.97    &     8.78   \\
50  & ATCDFS\_J033238.75-274632.5 &     0.072 $\pm$  0.030 & J033238.60-274631.4 & 0.23     &  2.4     &  21.533  &     \nodata         & \nodata  &    \nodata  &    \nodata \\
51a & ATCDFS\_J033239.46-275300.9 &     0.127 $\pm$  0.032 & J033239.47-275300.5 & 0.57     &  0.5     &  20.540  &     \nodata         & \nodata  &    \nodata  &    \nodata \\
51b &                             &                        & J033239.49-275301.6 & 0.43     &  0.7     &  20.400  &     \nodata         & \nodata  &    \nodata  &    \nodata \\
52a & ATCDFS\_J033239.64-274851.8 &     0.066 $\pm$  0.030 & J033239.67-274850.6 & 0.49     &  1.3     &  24.547  & J033239.68-274850.7 &  1.2     &     0.75    &     7.06   \\
52b &                             &                        & J033239.56-274851.7 & 0.47     &  1.2     &  22.552  &     \nodata         & \nodata  &    \nodata  &    \nodata \\
53  & ATCDFS\_J033240.82-275545.8 &     0.102 $\pm$  0.032 & J033240.84-275546.7 & 0.97     &  0.9     &  25.183  & J033240.84-275546.6 &  0.8     &     0.54    &     9.34   \\
54  & ATCDFS\_J033243.12-275514.2 &     0.070 $\pm$  0.034 & J033243.17-275514.7 & 0.79     &  0.7     &  23.991  &     \nodata         & \nodata  &    \nodata  &    \nodata \\
55a & ATCDFS\_J033244.16-275142.4 &     0.511 $\pm$  0.030 & J033244.05-275143.3 & \nodata  &  1.7     &  18.491  &     \nodata         & \nodata  &    \nodata  &    \nodata \\
55b &                             &                        & J033244.27-275141.1 & \nodata  &  1.9     &  19.166  & J033244.28-275141.0 &  2.1     &     0.51    & $<$ 0.33   \\
56  & ATCDFS\_J033244.93-274726.2 &     0.107 $\pm$  0.033 & J033244.86-274727.6 & 0.93     &  1.6     &  18.456  &     \nodata         & \nodata  &    \nodata  &    \nodata \\
57  & ATCDFS\_J033245.03-275438.8 &     0.122 $\pm$  0.038 & J033245.02-275439.6 & 0.99     &  0.7     &  18.987  & J033244.98-275438.7 &  0.7     &     0.14    & $<$ 0.60   \\
58  & ATCDFS\_J033246.03-275318.2 &     0.219 $\pm$  0.046 & \nodata             & \nodata  & \nodata  & $>$28.2  &     \nodata         & \nodata  &    \nodata  &    \nodata \\
59  & ATCDFS\_J033246.33-275328.6 &     0.080 $\pm$  0.038 & J033246.33-275327.0 & 0.76     &  1.7     &  23.639  &     \nodata         & \nodata  &    \nodata  &    \nodata \\
60a & ATCDFS\_J033246.78-275120.1 &     0.079 $\pm$  0.030 & J033246.84-275121.2 & 0.51     &  1.3     &  23.562  &     \nodata         & \nodata  &    \nodata  &    \nodata \\
60b &                             &                        & J033246.79-275118.8 & 0.45     &  1.4     &  24.769  &     \nodata         & \nodata  &    \nodata  &    \nodata \\
61  & ATCDFS\_J033248.02-275414.7 &     0.103 $\pm$  0.047 & J033248.05-275412.4 & 0.59     &  2.4     &  26.566  &     \nodata         & \nodata  &    \nodata  &    \nodata \\
62  & ATCDFS\_J033248.62-274934.9 &     0.097 $\pm$  0.030 & J033248.57-274934.3 & 0.92     &  1.0     &  23.425  &       (G578)\tablenotemark{c} &  0.5     &     0.09    & $<$ 0.36   \\
63  & ATCDFS\_J033300.91-275049.4 &     0.078 $\pm$  0.036 & J033300.98-275053.7 & 0.60     &  4.3     &  21.675  &     \nodata         & \nodata  &    \nodata  &    \nodata \\
64  & ATCDFS\_J033302.99-275147.3 &     0.080 $\pm$  0.036 & J033303.05-275145.8 & 0.89     &  1.7     &  25.134  & J033302.97-275146.4 &  1.0     &     0.69    &     0.93   \\
\enddata
\tablenotetext{a}{Optical sources deblended using SExtractor}
\tablenotetext{b}{Optical source not in the released ACS catalogue. The $z_{850}$ magnitude results from aperture photometry.}
\tablenotetext{c}{X-ray sources from Giacconi et al. (2002); the source is denoted by its number in the catalogue therein, while the flux in the 2-8\,keV band is an estimate based on the reported flux in the 2-10\,keV band and assuming a power-law photon index of 1.4.}
\end{deluxetable}

\clearpage

\begin{deluxetable}{rcccccccrrrrcc}
\tabletypesize{\tiny}
\rotate
\tablenum{2}
\tablecolumns{14}
\tablewidth{0pc}
\tablecaption{Characterization of faint radio sources \label{tab:info}}
\tablehead{
\colhead{No} & \colhead{$z_{sp}$} & \colhead{$Q_z$} & 
  \colhead{$z_{ph}$} & \colhead{type} &
  \colhead{$z_{ph}$\,/\,type\,/\, number of bands} &
  \colhead{$L_{\rm 1.4\,GHz}$} & \colhead{$L_X$} &
  \colhead{$f_{1.4}/f_{opt}$} &
  \colhead{$f_x/f_{opt}$} &
  \colhead{$HR$} &
  \colhead{$\Gamma$} & \colhead{class} &
  \colhead{class} \\
\colhead{} & \colhead{} & \colhead{} & 
\multicolumn{2}{c}{\hspace*{0.2truecm}(COMBO-17)} &
  \colhead{(Mobasher et al. 2004)} & 
  \colhead{(W\,Hz$^{-1}$)} & \colhead{(erg\,s$^{-1}$)} &
  \colhead{} &
  \colhead{} &
  \colhead{} & 
  \colhead{} & \colhead{(X-rays or radio)} & 
  \colhead{(optical)}
}
\startdata
 1  & \nodata                    & \nodata  & \nodata         & \nodata         & \nodata                             & \nodata  & \nodata  & $>$ 3.66    &    \nodata  &    \nodata  &    \nodata  & \nodata  & \nodata   \\
 2  &  0.720\tablenotemark{s}  & 1.0      & 0.686$\pm$0.080 & Galaxy          & 1.29\,$(0.99-1.59)$\,/\,2.67\,/\,13 & 23.30    & 42.49    &     1.84    & $<$-0.10    & $>$   0.51  & $<$-0.27    & AGN2     & LEX       \\
 3  & \nodata                    & \nodata  & \nodata         & \nodata         & 0.43\,$(0.13-3.73)$\,/\,2.00\,/\, 5 & 22.73    & \nodata  &     2.50    &    \nodata  &    \nodata  &    \nodata  & \nodata  & \nodata   \\
 4  &  0.310\tablenotemark{s}  & 2.0      & 0.328$\pm$0.004 & Galaxy          & 0.31\,$(0.14-0.48)$\,/\,1.00\,/\,15 & 22.76    & 41.34    &     0.15    & $<$-2.50    & $>$   0.29  & $<$ 0.21    & AGN2     & LEX/Sey2  \\
 5  & \nodata                    & \nodata  & 0.347$\pm$0.026 & Galaxy          & 0.44\,$(0.25-0.63)$\,/\,2.00\,/\,13 & 22.47    & \nodata  &     0.35    &    \nodata  &    \nodata  &    \nodata  & \nodata  & \nodata   \\
 6  &  0.543\tablenotemark{s}  & 2.0      & 0.582$\pm$0.017 & QSO             & 0.26\,$(0.09-0.56)$\,/\,3.67\,/\,15 & 24.32    & 44.08    &     1.13    &     0.16    &      -0.50  &     1.73    & QSO1     & BLAGN     \\
 7  &  0.733\tablenotemark{s}  & 2.0      & 0.762$\pm$0.007 & Galaxy          & 0.71\,$(0.49-0.93)$\,/\,1.00\,/\,13 & 23.70    & 42.28    &     0.99    &    -1.07    & $<$  -0.52  & $>$ 1.75    & AGN1     & LEX       \\
 8  &  1.610\tablenotemark{m}  & 2.0      & 1.013$\pm$0.212 & Galaxy          & 1.32\,$(1.02-1.62)$\,/\,1.00\,/\,14 & 24.34    & 42.47    &     1.80    &    -0.76    & $<$  -0.05  &     1.40\tablenotemark{a} & AGNX     & LEX       \\
 9  &  1.615\tablenotemark{s}  & 2.0      & 1.595$\pm$0.010 & QSO             & 0.76\,$(0.53-1.10)$\,/\,3.67\,/\,18 & 25.63    & 44.32    &     2.87    &     0.86    &      -0.49  &     1.69    & QSO1     & BLAGN     \\
10  & \nodata                    & \nodata  & \nodata         & \nodata         & \nodata                             & \nodata  & \nodata  &     2.33    &     0.82    &      -0.40  &     1.50    & UNCL     & \nodata   \\
11  &  0.576\tablenotemark{m}  & 2.0      & 0.594$\pm$0.019 & Galaxy          & 0.94\,$(0.69-1.19)$\,/\,2.00\,/\,18 & 23.23    & \nodata  &     1.01    &    \nodata  &    \nodata  &    \nodata  & \nodata  & LEX       \\
12  & \nodata                    & \nodata  & \nodata         & \nodata         & \nodata                             & \nodata  & \nodata  & $>$ 4.86    &    \nodata  &    \nodata  &    \nodata  & \nodata  & \nodata   \\
13a &  0.605\tablenotemark{s}  & 2.0      & 0.609$\pm$0.008 & Galaxy          & 0.63\,$(0.42-0.84)$\,/\,2.33\,/\,14 & 23.24    & 43.21    &     0.42    &    -0.61    &       0.09  &     0.58    & AGN2     & HEX       \\
13b & \nodata                    & \nodata  & \nodata         & \nodata         & \nodata                             & \nodata  & \nodata  &     1.31    &    \nodata  &    \nodata  &    \nodata  & \nodata  & \nodata   \\
14  & \nodata                    & \nodata  & \nodata         & \nodata         & 1.66\,$(1.31-2.26)$\,/\,4.00\,/\,15 & 24.33    & \nodata  &     2.23    &    \nodata  &    \nodata  &    \nodata  & \nodata  & \nodata   \\
15  &  0.569\tablenotemark{s}  & 2.0      & 0.583$\pm$0.018 & Galaxy          & 0.59\,$(0.38-0.80)$\,/\,2.00\,/\,15 & 22.98    & 43.12    &     0.56    &     0.00    &      -0.52  &     1.77    & AGN1     & BLAGN     \\
16  &  1.097\tablenotemark{s}  & 2.0      & 0.843$\pm$0.047 & Galaxy          & 1.09\,$(0.82-1.36)$\,/\,1.67\,/\,14 & 23.55    & 43.28    &     0.95    &    -0.43    &       0.63  &    -0.58    & AGN2     & LEX       \\
17  &  0.734\tablenotemark{s}  & 2.0      & 0.759$\pm$0.012 & Galaxy          & 0.77\,$(0.54-1.00)$\,/\,1.00\,/\,17 & 23.95    & 41.96    &     0.81    &    -1.67    & $<$  -0.58  & $>$ 1.92    & GALA     & LEX       \\
18  & \nodata                    & \nodata  & \nodata         & \nodata         & 1.04\,$(0.77-1.46)$\,/\,3.67\,/\,17 & 24.01    & \nodata  &     3.25    &    \nodata  &    \nodata  &    \nodata  & \nodata  & \nodata   \\
19  & \nodata                    & \nodata  & 0.965$\pm$0.014 & Galaxy          & 0.92\,$(0.67-1.17)$\,/\,1.00\,/\,14 & 25.30    & 42.38    &     2.23    &    -1.59    &    (-0.12) \tablenotemark{b} &    \nodata  & AGN2     & \nodata   \\
20  & \nodata                    & \nodata  & 1.049$\pm$0.150 & Galaxy          & 1.07\,$(0.80-1.34)$\,/\,1.33\,/\,13 & 24.01    & \nodata  &     1.49    &    \nodata  &    \nodata  &    \nodata  & \nodata  & \nodata   \\
21  & \nodata                    & \nodata  & 0.965$\pm$0.014 & Galaxy          & 0.92\,$(0.67-1.17)$\,/\,1.00\,/\,14 & 25.39    & 42.38    &     2.32    &    -1.59    &    (-0.12) \tablenotemark{b} &    \nodata  & AGN2     & \nodata   \\
22  &  0.229\tablenotemark{s}  & 2.0      & 0.226$\pm$0.013 & Galaxy          & 0.24\,$(0.08-0.40)$\,/\,1.33\,/\,14 & 22.53    & 41.21    &     0.29    &    -2.04    & $<$  -0.15  & $>$ 1.00    & UNCL     & LEX/Sey2  \\
23  &  0.524\tablenotemark{m}  & 2.0      & 0.460$\pm$0.018 & Galaxy          & 0.54\,$(0.34-0.74)$\,/\,1.00\,/\,18 & 23.25    & 41.13    &     0.54    & $<$-2.13    &    \nodata  &    \nodata  & UNCL     & LEX       \\
24  & \nodata                    & \nodata  & \nodata         & \nodata         & 1.54\,$(1.21-1.87)$\,/\,2.00\,/\,12 & 24.05    & \nodata  &     2.03    &    \nodata  &    \nodata  &    \nodata  & \nodata  & \nodata   \\
25  & \nodata                    & \nodata  & \nodata         & \nodata         & \nodata                             & \nodata  & \nodata  & $>$ 3.96    &    \nodata  &    \nodata  &    \nodata  & \nodata  & \nodata   \\
26  &  0.737\tablenotemark{s}  & 2.0      & 0.736$\pm$0.015 & Galaxy          & 0.74\,$(0.51-0.97)$\,/\,2.00\,/\,18 & 23.33    & 41.72    &     0.54    & $<$-1.92    &    \nodata  &     1.40\tablenotemark{a} & UNCL     & LEX/SF    \\
27  &  0.123\tablenotemark{l}  & 2.0      & 0.120$\pm$0.010 & Galaxy          & 0.18\,$(0.03-0.33)$\,/\,2.00\,/\,15 & 21.52    & \nodata  &    -0.06    &    \nodata  &    \nodata  &    \nodata  & \nodata  & LEX       \\
28  &  1.090\tablenotemark{s}  & 2.0      & 1.143$\pm$0.065 & Galaxy          & 1.01\,$(0.75-1.27)$\,/\,1.33\,/\,14 & 23.64    & 42.46    &     0.82    & $<$-1.49    & $>$   0.69  & $<$-0.76    & AGN2     & ABS       \\
29  &  0.734\tablenotemark{s}  & 2.0      & 0.754$\pm$0.155 & QSO             & 0.07\,$(0.00-0.54)$\,/\,6.00\,/\,14 & 25.64    & 44.42    &     2.34    &     0.42    &      -0.53  &     1.79    & QSO1     & BLAGN     \\
30  &  1.097\tablenotemark{v}  & 2.0      & 1.119$\pm$0.039 & Galaxy          & 1.06\,$(0.79-1.33)$\,/\,2.00\,/\,18 & 23.63    & 41.58    &     0.89    &    -1.83    & $<$   0.33  &    \nodata  & UNCL     & HEX       \\
31  &  0.247\tablenotemark{c}  & 2.0      & 0.240$\pm$0.030 & Galaxy          & 0.47\,$(0.28-0.66)$\,/\,1.33\,/\,18 & 22.22    & \nodata  &     0.16    &    \nodata  &    \nodata  &    \nodata  & \nodata  & LEX       \\
32  & \nodata                    & \nodata  & 0.391$\pm$0.012 & Galaxy          & 0.45\,$(0.26-0.64)$\,/\,1.33\,/\, 9 & 22.85    & \nodata  &     0.29    &    \nodata  &    \nodata  &    \nodata  & \nodata  & \nodata   \\
33  & \nodata                    & \nodata  & \nodata         & \nodata         & \nodata                             & \nodata  & \nodata  & $>$ 5.29    &    \nodata  &    \nodata  &    \nodata  & \nodata  & \nodata   \\
34  &  0.738\tablenotemark{s}  & 2.0      & 0.710$\pm$0.030 & Galaxy          & 0.76\,$(0.53-0.99)$\,/\,1.33\,/\,17 & 23.72    & 42.08    &     1.22    &    -1.08    &      -0.67  &     2.18    & AGN1     & LEX       \\
35  & \nodata                    & \nodata  & \nodata         & \nodata         & \nodata                             & \nodata  & \nodata  & $>$ 4.84    &    \nodata  &    \nodata  &    \nodata  & \nodata  & \nodata   \\
36  &  0.076\tablenotemark{s}  & 2.0      & 0.075$\pm$0.010 & Uncl.           & 0.16\,$(0.01-0.31)$\,/\,2.33\,/\,18 & 22.18    & 40.20    &     0.06    &    -2.39    & $<$  -0.79  & $>$ 2.65    & GALA     & LEX       \\
37  &  0.953\tablenotemark{v}  & 2.0      & 0.811$\pm$0.022 & Galaxy          & 1.01\,$(0.75-1.27)$\,/\,2.33\,/\,18 & 23.67    & 41.72    &     0.90    & $<$-1.63    & $>$  -0.40  &    \nodata  & UNCL     & LEX       \\
38  &  0.076\tablenotemark{s}  & 2.0      & 0.086$\pm$0.012 & Galaxy          & 0.11\,$(0.00-0.26)$\,/\,2.00\,/\,18 & 21.80    & 40.23    &    -0.17    &    -2.39    &      -0.56  &     1.87    & GALA     & LEX       \\
39  & \nodata                    & \nodata  & \nodata         & \nodata         & \nodata                             & \nodata  & \nodata  &     2.83    &    \nodata  &    \nodata  &    \nodata  & \nodata  & \nodata   \\
40  & \nodata                    & \nodata  & \nodata         & \nodata         & \nodata                             & \nodata  & \nodata  &     2.94    &    \nodata  &    \nodata  &    \nodata  & \nodata  & \nodata   \\
41a &  2.223\tablenotemark{s}  & 2.0      & \nodata         & \nodata         & 2.21\,$(1.79-2.63)$\,/\,2.67\,/\,18 & 24.27    & 42.67    &     1.39    &    -0.95    & $<$   0.02  &     1.40\tablenotemark{a} & AGNX     & HEX       \\
41b & \nodata                    & \nodata  & 0.841$\pm$0.055 & Galaxy          & 0.78\,$(0.55-1.01)$\,/\,5.33\,/\,17 & 23.28    & \nodata  &     1.46    &    \nodata  &    \nodata  &    \nodata  & \nodata  & \nodata   \\
42  & \nodata                    & \nodata  & \nodata         & \nodata         & \nodata                             & \nodata  & \nodata  & $>$ 3.67    &    \nodata  &    \nodata  &    \nodata  & UNCL     & \nodata   \\
43  & \nodata                    & \nodata  & \nodata         & \nodata         & \nodata                             & \nodata  & \nodata  &     3.54    &    \nodata  &    \nodata  &    \nodata  & \nodata  & \nodata   \\
44  &  0.038\tablenotemark{s}  & 2.0      & 0.058$\pm$0.006 & Galaxy          & 0.28\,$(0.11-0.45)$\,/\,1.33\,/\,15 & 21.01    & 39.72    &    -0.79    & $<$-3.49    & $>$   0.31  & $<$ 0.17    & ULX      & LEX       \\
45  & \nodata                    & \nodata  & \nodata         & \nodata         & \nodata                             & \nodata  & \nodata  &     2.80    &    \nodata  &    \nodata  &    \nodata  & \nodata  & \nodata   \\
46  &  2.578\tablenotemark{s}  & 2.0      & \nodata         & \nodata         & 2.51\,$(2.05-2.97)$\,/\,2.00\,/\,14 & 24.48    & 42.62    &     2.16    &    -0.32    &       0.30  &     0.19    & AGN2     & HEX       \\
47a & \nodata                    & \nodata  & 0.528$\pm$0.031 & Galaxy          & 0.67\,$(0.45-0.89)$\,/\,2.00\,/\,15 & 22.89    & \nodata  &     0.87    &    \nodata  &    \nodata  &    \nodata  & \nodata  & \nodata   \\
47b & \nodata                    & \nodata  & 0.528$\pm$0.031 & Galaxy          & 0.67\,$(0.45-0.89)$\,/\,2.00\,/\,15 & 22.89    & \nodata  &     0.74    &    \nodata  &    \nodata  &    \nodata  & \nodata  & \nodata   \\
48a & \nodata                    & \nodata  & \nodata         & \nodata         & 1.36\,$(1.05-1.67)$\,/\,2.33\,/\,17 & 24.23    & \nodata  &     2.04    &    \nodata  &    \nodata  &    \nodata  & \nodata  & \nodata   \\
48b & \nodata                    & \nodata  & \nodata         & \nodata         & 1.36\,$(1.05-1.67)$\,/\,2.33\,/\,17 & 24.23    & \nodata  &     2.02    &    \nodata  &    \nodata  &    \nodata  & \nodata  & \nodata   \\
49  &  1.603\tablenotemark{s}  & 2.0      & 1.015$\pm$0.127 & Galaxy          & 1.76\,$(1.40-2.12)$\,/\,1.67\,/\,14 & 24.10    & 44.30    &     1.82    &     1.31    &      -0.51  &     1.75    & QSO1     & HEX       \\
50  & \nodata                    & \nodata  & 0.567$\pm$0.032 & Galaxy          & 0.58\,$(0.37-0.79)$\,/\,2.67\,/\,18 & 22.93    & \nodata  &     0.91    &    \nodata  &    \nodata  &    \nodata  & \nodata  & \nodata   \\
51a & \nodata                    & \nodata  & 0.652$\pm$0.012 & Galaxy          & 0.61\,$(0.40-0.82)$\,/\,1.00\,/\,15 & 23.32    & \nodata  &     0.76    &    \nodata  &    \nodata  &    \nodata  & \nodata  & \nodata   \\
51b & \nodata                    & \nodata  & 0.650$\pm$0.014 & Galaxy          & 0.61\,$(0.40-0.82)$\,/\,1.00\,/\,15 & 23.32    & \nodata  &     0.70    &    \nodata  &    \nodata  &    \nodata  & \nodata  & \nodata   \\
52a &  3.064\tablenotemark{s}  & 2.0      & \nodata         & \nodata         & 1.17\,$(0.89-1.45)$\,/\,2.33\,/\,13 & 24.61    & 43.83    &     2.08    &     0.81    &       0.18  &     0.42    & AGN2     & HEX       \\
52b & \nodata                    & \nodata  & 1.400$\pm$0.076 & Galaxy          & 0.85\,$(0.61-1.09)$\,/\,1.33\,/\,14 & 23.82    & \nodata  &     1.28    &    \nodata  &    \nodata  &    \nodata  & \nodata  & \nodata   \\
53  &  0.625\tablenotemark{s}  & 0.5      & \nodata         & \nodata         & 1.81\,$(1.44-2.20)$\,/\,2.00\,/\,11 & 23.18    & 42.76    &     2.52    &     0.93    &       0.40  &    -0.02    & AGN2     & ABS       \\
54  & \nodata                    & \nodata  & \nodata         & \nodata         & 1.63\,$(1.28-1.99)$\,/\,2.67\,/\,14 & 24.00    & \nodata  &     1.88    &    \nodata  &    \nodata  &    \nodata  & \nodata  & \nodata   \\
55a &  0.279\tablenotemark{s}  & 1.0      & 0.264$\pm$0.022 & Galaxy          & 0.46\,$(0.27-0.65)$\,/\,1.33\,/\,15 & 23.07    & \nodata  &     0.54    &    \nodata  &    \nodata  &    \nodata  & \nodata  & ABS       \\
55b &  0.279\tablenotemark{s}  & 2.0      & 0.262$\pm$0.049 & Galaxy          & 0.28\,$(0.11-0.45)$\,/\,1.33\,/\,15 & 23.07    & 41.33    &     0.82    &    -1.51    & $<$  -0.72  & $>$ 2.35    & GALA     & HEX/Sey2  \\
56  &  0.214\tablenotemark{s}  & 2.0      & 0.187$\pm$0.031 & Galaxy          & 0.37\,$(0.19-0.55)$\,/\,2.33\,/\,16 & 22.14    & \nodata  &    -0.15    &    \nodata  &    \nodata  &    \nodata  & \nodata  & LEX       \\
57  &  0.458\tablenotemark{s}  & 2.0      & 0.467$\pm$0.008 & Galaxy          & 0.45\,$(0.26-0.64)$\,/\,2.33\,/\,15 & 22.94    & 41.38    &     0.12    &    -2.14    & $<$  -0.09  &     1.40\tablenotemark{a} & UNCL     & LEX       \\
58  & \nodata                    & \nodata  & \nodata         & \nodata         & \nodata                             & \nodata  & \nodata  & $>$ 4.06    &    \nodata  &    \nodata  &    \nodata  & \nodata  & \nodata   \\
59  & \nodata                    & \nodata  & 0.833$\pm$0.054 & Galaxy          & 1.40\,$(1.08-1.72)$\,/\,1.67\,/\,11 & 23.37    & \nodata  &     1.80    &    \nodata  &    \nodata  &    \nodata  & \nodata  & \nodata   \\
60a & \nodata                    & \nodata  & 1.065$\pm$0.106 & Galaxy          & 1.25\,$(0.89-1.55)$\,/\,3.67\,/\,15 & 23.62    & \nodata  &     1.76    &    \nodata  &    \nodata  &    \nodata  & \nodata  & \nodata   \\
60b & \nodata                    & \nodata  & \nodata         & \nodata         & 0.17\,$(0.02-3.10)$\,/\,3.67\,/\,13 & 21.79    & \nodata  &     2.25    &    \nodata  &    \nodata  &    \nodata  & \nodata  & \nodata   \\
61  & \nodata                    & \nodata  & \nodata         & \nodata         & \nodata                             & \nodata  & \nodata  &     3.08    &    \nodata  &    \nodata  &    \nodata  & \nodata  & \nodata   \\
62  &  1.117\tablenotemark{l}  & 0.5      & 0.987$\pm$0.105 & Galaxy          & 1.06\,$(0.79-1.33)$\,/\,1.00\,/\, 9 & 23.76    & \nodata  &     1.80    &    -0.56    &    \nodata  &    \nodata  & UNCL     & \nodata   \\
63  & \nodata                    & \nodata  & 0.816$\pm$0.059 & Galaxy          & 0.87\,$(0.62-1.12)$\,/\,2.00\,/\,13 & 23.34    & \nodata  &     1.00    &    \nodata  &    \nodata  &    \nodata  & \nodata  & \nodata   \\
64  & \nodata                    & \nodata  & \nodata         & \nodata         & 0.45\,$(0.25-1.35)$\,/\,2.00\,/\,10 & 22.74    & 42.05    &     2.40    &     1.01    &      -0.55  &     1.81    & AGN1     & \nodata   \\
\enddata
\tablenotetext{a}{Generic photon index due to low number of counts}
\tablenotetext{b}{Hardness Ratio as given in Giacconi et al. (2002), calculated using the 0.5-2 and 2-10\,keV bands}
\tablenotetext{s}{Spectra from Szokoly et al. (2004)}
\tablenotetext{v}{Spectra from Vanzella et al. (2004)}
\tablenotetext{l}{Spectra from Le F\`{e}vre et al. (2004)}
\tablenotetext{c}{Spectra from Croom, Warren, \& Glazebrook (2001)}
\tablenotetext{m}{Spectra from Mignoli et al. (2005)}
\end{deluxetable}

\clearpage

\pagestyle{plaintop}

\begin{figure}
\begin{center}
\rotatebox{0}{
\includegraphics[width=450pt]{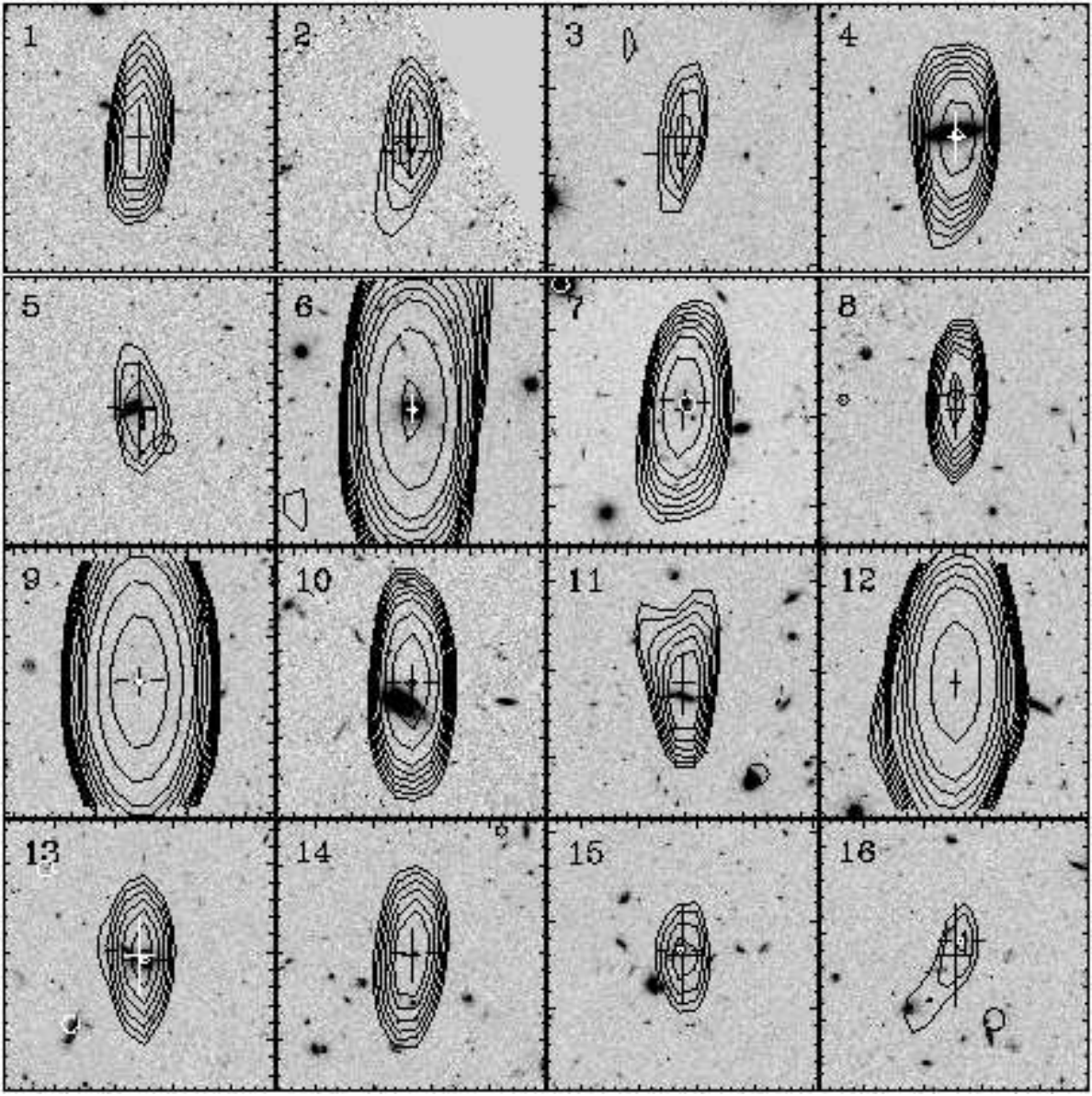}
}

\caption{Optical $B_{435} + V_{606} + i_{775} + z_{850}$ images for radio 
sources in the GOODS-S ACS field with radio contours overlayed and X-ray 
detections indicated. Source number from Table~\ref{tab:ids} is printed in 
the upper left corner of each image, which is roughly 30\arcsec~wide, 
centered on the radio source coordinates and with North being up and East 
to the right. The crosses indicate the 3\,$\sigma$ radio position error 
regions. Radio contours go from 45 to 93\,$\mu$Jy in factors of 1.2, from 
110 to 302\,$\mu$Jy in factors of 1.4, and then from 420\,$\mu$Jy to 13\,mJy 
in factors of 2. The horizontal lines enclose the assumed optical 
counterparts of radio sources, while circles indicate X-ray detections with
sizes that denote the X-ray position error. For sources 
\#19, \#21, \#42 and \#62, the double circles indicate X-ray detections 
present only in the \citet{Giacconi02} catalog.
\label{fig:pstamps}}
\end{center}
\end{figure}

\addtocounter{figure}{-1}

\begin{figure}
\begin{center}
\rotatebox{0}{
\includegraphics[width=450pt]{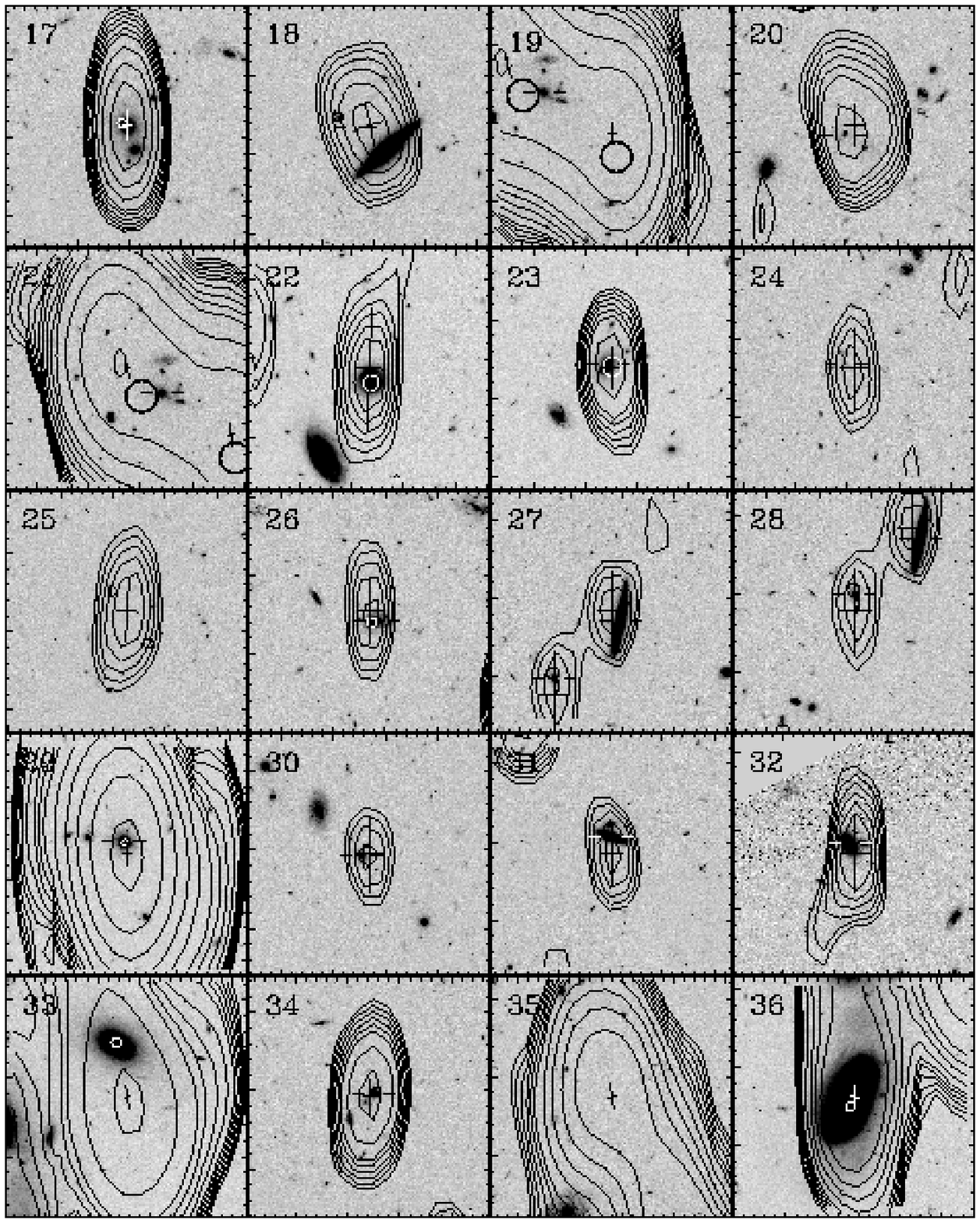}
}

\caption{cont.
\label{fig:pstamps2}}
\end{center}
\end{figure}

\addtocounter{figure}{-1}

\begin{figure}
\begin{center}
\rotatebox{0}{
\includegraphics[width=450pt]{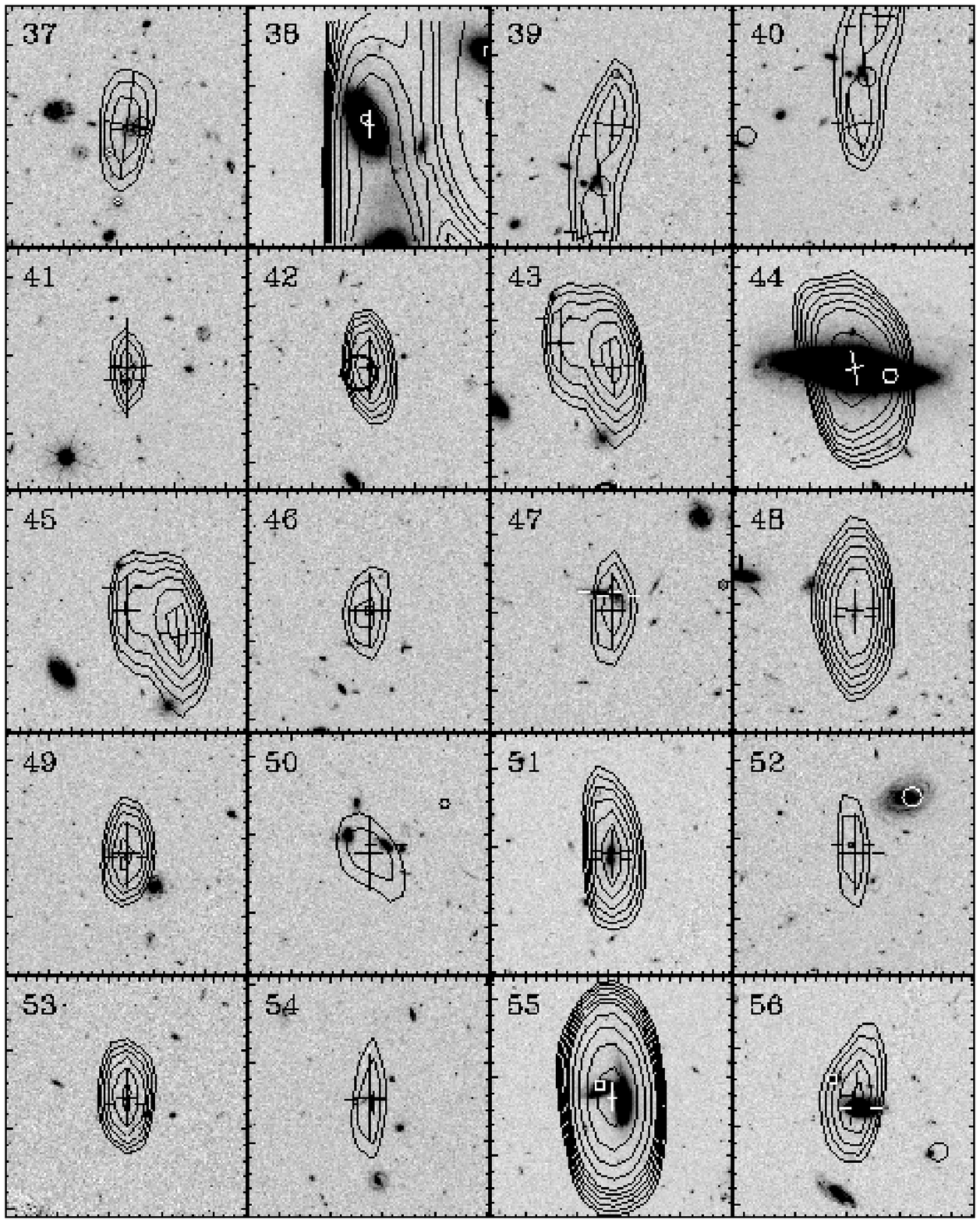}
}

\caption{cont.
\label{fig:pstamps3}}
\end{center}
\end{figure}

\addtocounter{figure}{-1}

\begin{figure}
\begin{center}
\rotatebox{0}{
\includegraphics[width=450pt]{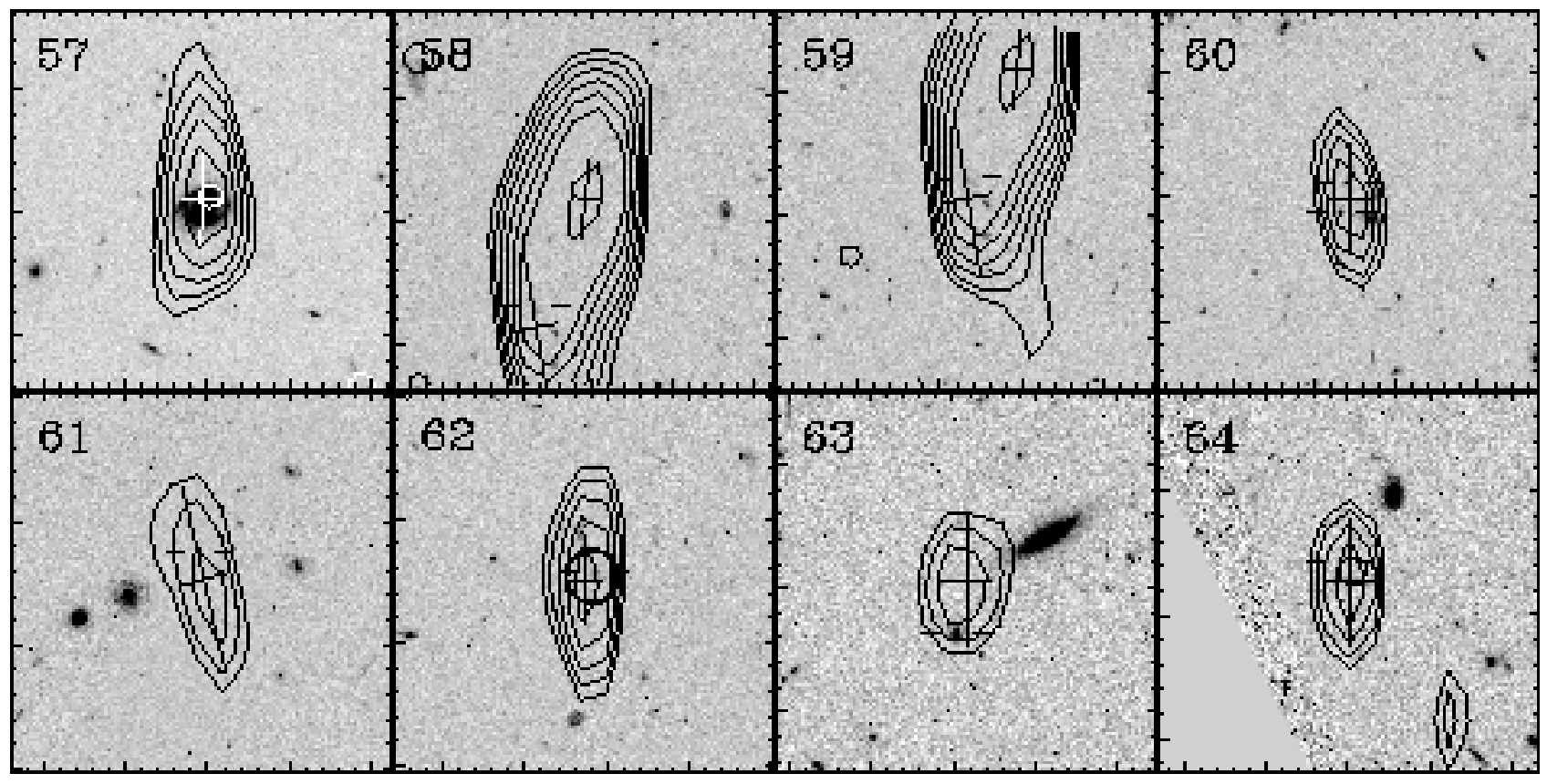}
}

\caption{cont.
\label{fig:pstamps4}}
\end{center}
\end{figure}

\pagebreak

\end{document}